\pdfoutput=1

\documentclass[12pt]{article}

\usepackage{amsmath,amsthm,amssymb,amsfonts}
\usepackage{graphicx,psfrag,epsf}
\usepackage{enumerate}
\usepackage{natbib}
\usepackage[english]{babel}
\usepackage{url} % not crucial - just used below for the URL 

\newcommand{\blind}{1}

\usepackage{bbm}
\usepackage[dvipsnames]{xcolor}
\usepackage{titlesec}
\usepackage[labelfont={bf,it},textfont=it]{caption}
\usepackage{tabstackengine}
\titlelabel{\thetitle.\quad}
\DeclareMathOperator*{\argmin}{arg\,min}
\DeclareMathOperator*{\argmax}{arg\,max}
\renewcommand{\thefigure}{\arabic{section}.\arabic{figure}}
\newcommand\mysup[3][3pt]{%
 \def\stackalignment{c}\ensurestackMath{\stackunder[#1]{\sup #2}{%
  \makebox[0pt]{$\scriptstyle #3$}}}%
}
\newcommand\var{{\rm Var}}

\begin{document}

\def\d{{\rm d}}
\def\P{{\rm P}}
\def\mX{{\mathcal{X}}}
\def\bX{{\mathbf{X}}}
\def\bhB{{\mathbf{\widehat{B}}}}
\def\bone{{\boldsymbol{1}}}
\def\bhT{{\mathbf{\widehat{\Theta}}}}
\def\bLambda{{\bf \Lambda}}
\def\L{{\cal L}}
\def\F{{\cal F}}
\def\cT{{\cal T}}
\def\dif{{\rm d}}
\def\lc{\left\lceil}   
\def\rc{\right\rceil}
\newtheorem{definition}{\bf Definition}[section]
\newtheorem{theorem}{\bf Theorem}[section]
\newtheorem{proposition}{\bf Proposition}[section]
\newtheorem{corollary}{\bf Corollary}[section]
\newtheorem{remark}{\bf Remark}

\def\spacingset#1{\renewcommand{\baselinestretch}%
{#1}\small\normalsize} \spacingset{1}

%%%%%%%%%%%%%%%%%%%%%%%%%%%%%%%%%%%%%%%%%%%%%%%%%%%%%%%%%%%%%%%%%%%%%%%%%%%%%%

\if1\blind
{
  \title{\bf Multiresolution analysis of point processes and statistical thresholding for wavelet-based intensity estimation}
  \author{Youssef Taleb\thanks{ 
  \href{mailto:youssef.taleb12@imperial.ac.uk}{youssef.taleb12@imperial.ac.uk}.
    The authors gratefully acknowledge \textit{EPSRC}}\hspace{.2cm}\\  
    and \\
   Edward A. K. Cohen \\
    Department of Mathematics, Imperial College London, \\ South Kensington Campus, London, SW7 2AZ, United Kingdom.}
\date{}
  \maketitle
} \fi

\if0\blind
{
  \bigskip
  \bigskip
  \bigskip
  \begin{center}
    {\LARGE\bf Wavelet analysis of first-order properties of point processes}
\end{center}
  \medskip
} \fi

\bigskip
\begin{abstract}
%Since their introduction in the 1980s, wavelets have played a major role in the multiresolution analysis of signals. Considering point processes, they have been successfully used to estimate their intensity in a non-parametric fashion. Over the last twenty years both linear and non-linear wavelet estimators have been proposed in an accuracy seeking context. However, the theory resulting from the intersection of point processes and wavelet analysis has not been fully explored, especially a sequential multiresolution analysis for point processes is lacking for further applications. We address this with newly defined multiscale properties named Jth-level homogeneity and Lth-level innovation, which together form a first-order multiresolution analysis of Poisson processes. We then propose a natural application of this framework with statistical thresholding models for the intensity. Our thresholding procedures are based on likelihood ratio tests designed to verify these two properties when the intensity is reconstructed with Haar wavelets. Their expected Root Mean Square Error (RMSE) values are eventually compared on three intensity models under different scenarios.

We take a wavelet based approach to the analysis of point processes and the estimation of the first order intensity under a continuous time setting. A multiresolution analysis of a point process is formulated which motivates the definition of homogeneity at different scales of resolution, termed $J$-th level homogeneity. Further to this, the activity in a point processes' first order behavior at different scales of resolution is also defined and termed $L$-th level innovation. Likelihood ratio tests for both these properties are proposed with asymptotic distributions provided, even when only a single realization of the point process is observed. The test for $L$-th level innovation forms the basis for a collection of statistical strategies for thresholding coefficients in a wavelet based estimator of the intensity function. These thresholding strategies are shown to outperform the existing local hard thresholding strategy on a range of simulation scenarios.

\end{abstract}

\noindent%
\vfill

\newpage
\spacingset{1.45} % DON'T change the spacing!
 \section{Introduction}
\label{sec:intro}
The development of wavelet theory has been one of the most significant advances in signal and image processing. Wavelets' ability to decompose an object at different scales makes them ideal for understanding underlying structures in random processes. Based on their success in analyzing time series \citep{Percival2000}, there has been an ever increasing interest in applying wavelets to point processes \citep[e.g.][]{brillinger1997some, Cohen2014}. Representing a point process as $N(A)$, a random integer indicating the number of events that have occurred in the set $A\subset\mathbb{R}$, one may use the notation $N(t)$ to be equal to $N((0,t])$ for $t>0$, $-N((t,0])$ for $t<0$ and $N(0)=0$ \citep{Daley1998}. Wavelets have most commonly been used to estimate the first order intensity (rate) function $\lambda:\mathbb{R}\rightarrow\mathbb{R}_{\geq 0}$ defined as $\lambda(t) = E\{\dif N(t)\}/\dif t$. Here, $\dif N(t)$ denotes the differential process $N(t+\dif t)-N(t)$. This is based on the fact we can represent any $L^2(\mathbb{R})$ function as a linear combination of basis functions. Namely, for some $j_0\in\mathbb{Z}$ and father and mother wavelet pair $(\phi,\psi)$,
 \begin{equation}
 \label{lambda}
 \lambda(t) = \sum\limits_{k \in \mathbb{Z}}\alpha_{j_{0},k}\phi_{j_{0},k}(t)+\sum\limits_{j\geq j_0}\sum\limits_{k \in \mathbb{Z}}\beta_{j,k}\psi_{j,k}(t)
 \end{equation}
 where $\phi_{j_0,k} (x) = 2^{j_0/2}\phi(2^{j_0}x-k)$ and $\psi_{j,k} (x) = 2^{j/2}\psi(2^{j}x-k)$, provided $\lambda \in L^2(\mathbb{R})$.  To estimate $\lambda$, the task becomes estimating the coefficients $\{\alpha_{j_0,k} \equiv \langle\lambda,\phi_{j_0,k}\rangle;k\in\mathbb{Z}\}$ and $\{\beta_{j,k}\equiv \langle\lambda,\psi_{j,k}\rangle;j\geq j_0, k\in\mathbb{Z}\}$, where $\langle f_{1},f_{2}\rangle = \int_\mathbb{R} f_1(t)f_2^\ast(t)\dif t$ is the usual inner product on $L^{2}(\mathbb{R})$. This can be achieved by computing the stochastic integrals $\widehat{\alpha}_{j_0,k} = \int_{\mathbb{R}} \phi_{j_0,k}(t)\dif N(t) = \sum_{\tau_i\in\mathcal{E}}\phi_{j_0,k}(\tau_i)$ and $\widehat{\beta}_{j,k} = \int_{\mathbb{R}} \psi_{j,k}(t)\dif N(t) = \sum_{\tau_i\in\mathcal{E}}\psi_{j,k}(\tau_i)$, where $\mathcal{E}$ is the set of random event times of the process. Both $\widehat{\alpha}_{j_0,k}$ and $\widehat{\beta}_{j,k}$ can easily be shown to be unbiased estimators of $\alpha_{j_0,k}$ and $\beta_{j,k}$, respectively. Restricting the wavelet reconstruction up to some maximum resolution $J\geq j_0$ in \eqref{lambda},  one can construct the estimator 
 \begin{equation}
 \label{lambdahat}
 \widehat\lambda^J(t) = \sum\limits_{k \in \mathbb{Z}}\widehat\alpha_{j_{0},k}\phi_{j_{0},k}(t)+\sum\limits_{j= j_0}^J\sum\limits_{k \in \mathbb{Z}}\widehat\beta_{j,k}\psi_{j,k}(t)
 \end{equation}
 which is asymptotically unbiased as $J \rightarrow \infty$ under standard regularity assumptions on $N$ \citep{DeMiranda2011}. As in the classical wavelet regression setting \citep{Donoho1993}, or when using wavelets to estimate probability density functions \citep{Hardle1998}, it is then typical that shrinkage or thresholding procedures are applied to the coefficients to reduce the variance of the estimator $\widehat\lambda^J$.
 
Estimating the intensity of a point process has of course been addressed numerous times in either parametric \citep[e.g.][]{Rathbun1994} or non-parametric methods \citep[e.g.][]{brillinger1975,Aalen1978, Ramlau1983,Patil2004}. In the specific case of wavelet based estimation, a non-parametric method, the approaches can be split into discrete-time and continuous-time methods. Discrete time methods \citep[e.g.][]{timmermann1999multiscale,kolaczyk1999wavelet,Kolaczyk2000,Fryzlewicz2004} typically apply a discrete wavelet transform (DWT) to the aggregated process $\{N_t;t\in\mathbb{Z}\}$, where $N_t \equiv N(t+1)-N(t)$ and then perform a shinkage procedure. \cite{Besbeas2004} offers a comprehensive review of discrete time methods and provides a simulation study comparing various thresholding schemes. 
 
Under the continuous time framework, the setting of this paper, \cite{brillinger1997some} proposes the estimator in (\ref{lambdahat}), as well as an estimator for the second-order intensity. The shrinkage procedure $\widehat{\beta}_{jk}\rightarrow w(\widehat{\beta}_{j,k}/s_{j,k})$ is proposed where $s_{j,k}$ is an estimate of the standard error in $\widehat{\beta}_{j,k}$ and $w(u) = (1-u^{-2})_+$ is the Tukey function. Although applied to California earthquake data, the properties of the estimator are not studied in any detail. De Miranda \citeyearpar{DeMiranda2008} offers the first proper treatment of the continuous time formulation, providing the characteristic and density functions for the estimators of the coefficients $\{\alpha_{j_0,k};k\in\mathbb{Z}\}$ and $\{\beta_{j,k};j\geq j_0,k\in\mathbb{Z}\}$ in terms of the basis  $(\phi,\psi)$  for any continuous compactly supported wavelet of known closed form. This result is theoretically interesting but cannot be readily exploited as, apart from the Haar family, wavelets that fulfil all these criteria are rare and exotic. This work is extended in \cite{DeMiranda2011} to provide first and second order moments for the linear (no thresholding) intensity estimator for any compactly supported wavelet of known closed form. With $\mathbbm{1}_A(x)$ representing the characteristic function of the set $A$, they also propose a hard threshold $\widehat{\beta}_{j,k}\rightarrow \widehat{\beta}_{j,k}(1-\mathbbm{1}_{[-\omega s_{j,k},\omega s_{j,k}]}(\widehat{\beta}_{j,k}))$ ($\omega$ typically set to 3) but it is given little treatment.  
  
Further thresholding procedures have been proposed in \cite{Bigot2013} under a Meyer wavelet basis and in \cite{Reynaud-Bouret2010} under any biorthogonal wavelet basis. Both of these estimators are shown to achieve near optimal performance in the asymptotic setting that $M$, the number of observed independent realizations of the point process, goes to infinity. Further, the thresholding procedure of \cite{Reynaud-Bouret2010} does not require a compactly supported and bounded intensity to achieve asymptotic optimality. However, both thresholds are proportional to $\log(M)$ and are therefore only non-zero when $M>1$, a highly restrictive condition for application purposes where one may only ever be able to observe a single realization. A thresholding procedure that can be applied in the $M=1$ setting but for which the statistical properties are still tractable is therefore clearly desirable. In this paper, we consider a wavelet based multiresolution analysis of a point process to propose  statistical thresholding procedures of the intensity function. Statistical thresholding has previously been considered in \cite{abramovich1995thresholding} in the classical wavelet regression setting. Here we adapt it for point processes and show it is capable of providing estimates with just a single realization of the process ($M=1$), while being grounded in a statistically principled and tractable framework. 

In Section \ref{sec:math} we provide a background to wavelet estimation of point process intensities. We extend existing results to show that the linear wavelet estimator of $\lambda$ has a scaled Poisson distribution under a Poisson process and the Haar wavelet basis . Then in Section \ref{sec:test} we develop the theoretical framework for a wavelet-based multiresolution analysis of a point process. Considering the first order properties of a point process to be due to activity on different scales, under the Haar basis we define different levels of homogeneity, which we term $J$-th level homogeneity in reference to the particular scale $J$ at which we are analyzing the point process. We provide a likelihood ratio test (LRT) for these different levels of homogeneity for the class of Poisson processes, providing the asymptotic distribution for the LRT statistic under the null hypothesis.  We then consider a more general test for whether the intensity function exhibits activity at a particular scale, which we term $L$-th level innovation. Again, we provide a LRT for this property for the class of Poisson processes under the Haar wavelet basis . 

In Section \ref{sec:thresh}, we demonstrate how the LRT for $L$-th level innovation can be used as a method of statistical thresholding for wavelet coefficients, for which we propose three different forms: local, intermediate and global. Importantly, we demonstrate that under our LRT framework increasing $M$ and increasing the intensity of the process are equivalent to one another, and hence indistinguishable in the asymptotic analysis. We are therefore able to use the asymptotic distributions to draw reliable inference and threshold the intensity in the $M=1$ setting. We finish by providing a comprehensive simulation study comparing the three different statistical thresholding procedures presented in this paper with the hard thresholding procedure given in \cite{DeMiranda2011}. We demonstrate that one or more of the proposed statistical thresholding procedures outperform this hard thresholding in almost all circumstances.

A discussion on how the estimation and statistical thresholding procedures presented in this paper can be extended to Daubechies D4 wavelets can be found in Appendix \ref{D2Q}. Further discussions on the LRTs, including boundary cases can be found in Appendix \ref{LRTdiscussion}, all proofs are provided in Appendix \ref{proofs}, and results of a comprehensive simulation study can be found in Appendix \ref{apsec:sims}.

\section{Wavelets and Estimation of the Intensity}
\label{sec:math}
\setcounter{figure}{0}  

In this section we provide a brief background to wavelet estimation of point process intensities. We will restrict ourselves to simple point processes, i.e. point processes that satisfy $N(\{t\}) \in \{0,1\}$  almost surely for all $t\in\mathbb{R}$.
 
\subsection{Wavelets and multiresolution analysis}
\label{wavelet}
We summarize here essential definitions and results on wavelets that need to be stated prior to their application to the intensity function. The theory presented here follows the work of \cite{Meyer1992}. 
\begin{definition}
A \textbf{multiresolution approximation} of  $L^{2}(\mathbb{R}^{n})$ is an increasing sequence $V_{j}$, $j \in \mathbb{Z}$, of closed linear subspaces of $L^{2}(\mathbb{R}^{n})$ with the following properties:
\begin{enumerate}
\item $\bigcap\limits^{\infty}_{j=-\infty} V_{j} = \{0\}, \quad \bigcup\limits^{\infty}_{j=-\infty} V_{j}$ is dense in  $L^{2}(\mathbb{R}^{n})$;
\item for all $f \in L^{2}(\mathbb{R}^{n})$ and $j \in \mathbb{Z}$, $f(\cdot) \in V_{j} \iff f(2\cdot) \in V_{j+1}$;
\item for all $f \in L^{2}(\mathbb{R}^{n})$ and $k \in \mathbb{Z}^{n}$, $f(\cdot) \in V_{0} \iff f(\cdot-k) \in V_{0}$;
\item there exists a function $g \in V_{0}$, such that the sequence $g(\cdot-k), k \in \mathbb{Z}^{n}$, is a Riesz basis of the space $V_{0}$.
\end{enumerate}
\label{def_mra}
\end{definition}
It is also shown in \cite{Meyer1992} that for a Riesz basis $g(\cdot -k), k \in \mathbb{Z}^{n}$ of $V_{0}$, the sequence $\phi(\cdot -k), k \in \mathbb{Z}^{n}$ defined by $\Phi(\xi) = G(\xi)(\sum\limits_{k \in \mathbb{Z}^{n}}|G(\xi+2k\pi)|^{2})^{-1/2}$ is the canonical orthonormal basis of $V_{0}$, where $\Phi$ and $G$ are the Fourier transforms of $\phi$ and $g$, respectively. $\phi$ is called either the father wavelet or scaling function. In this paper, we are concerned with point processes on the real line, and therefore we focus on the space $L^{2}(\mathbb{R})$. Defining $W_{j}$ to be the orthogonal complement of $V_{j}$ in $V_{j+1}$, Definition \ref{def_mra} allows us to write \begin{equation}\label{MRA}L^{2}(\mathbb{R}) =\overline{ V_{j_{0}} \oplus  \bigoplus\limits_{j = j_{0}}^{\infty}W_{j}}\qquad \text{or} \qquad L^{2}(\mathbb{R}) = \overline{\bigoplus\limits_{j = -\infty}^{\infty}W_{j}}.\end{equation} The spaces $V_j$ each have the basis $\{\phi_{j,k}(x) := 2^{j/2}\phi(2^j x-k),k\in\mathbb{Z}\}$ and are called the approximation spaces. The spaces $W_{j}$ are called detail spaces and each have the orthonormal basis $\{\psi_{j,k}(x) := 2^{j/2}\psi(2^jx-k),k\in\mathbb{Z}\}$, where $\psi(x)$ is called the mother wavelet and is constructed from the father wavelet. The mappings $f(\cdot)\rightarrow 2^{j/2}f(2^j\cdot - k)$ are called dyadic transformations. Consequently, a fundamental result from (\ref{MRA})  is that for any $j_0\in \mathbb{Z}$, the set $\left\{\phi_{j_0,k}; k\in\mathbb{Z}\right\}\cup\left\{\psi_{j,k}; j \geq j_0, k \in \mathbb{Z} \right\}$ forms an orthonormal basis for $L^{2}(\mathbb{R})$. Furthermore, for any $j_0 \in \mathbb{Z}$ a function $ f \in L^{2}(\mathbb{R})$ can be decomposed as 
 \begin{equation}
f(x) = \sum\limits_{k \in \mathbb{Z}}\langle f,\phi_{j_0,k}\rangle\phi_{j_0,k}(x) + \sum\limits_{j\geq j_0}\sum\limits_{k \in \mathbb{Z}}\langle f,\psi_{j,k}\rangle\psi_{j,k}(x). \label{wav_exp_f}
\end{equation} 
This identity, which illustrates the idea of multiscale analysis, will be used to decompose the first order intensity of a point process. In practice, a function $f \in L^{2}(\mathbb{R})$ is often approximated by its projection onto a specific approximation space $V_J = V_{j_0} \oplus  \bigoplus\limits_{j = j_0}^{J-1}W_{j}$, with $J > j_0$. Expansion (\ref{wav_exp_f}) is then reduced to:
\begin{equation}
f^J(x)=   \sum\limits_{k \in \mathbb{Z}}\langle f,\phi_{J,k}\rangle\phi_{J,k}(x)\label{wav_exp_f_J} = \sum\limits_{k \in \mathbb{Z}}\langle f,\phi_{j_0,k}\rangle\phi_{j_0,k}(x) + \sum\limits_{j = j_0}^{J-1}\sum\limits_{k \in \mathbb{Z}}\langle f,\psi_{j,k}\rangle\psi_{j,k}(x) .
\end{equation} 
As we increase $J$, the function $f^J\in  V_J$ approximates $f$ with ever increasing accuracy such that $\|f^J - f\|_2\rightarrow 0$ as $J\rightarrow\infty$, where $\|\cdot\|_2 = \sqrt{\langle \cdot,\cdot\rangle}$ is the $L^2$ norm. 
\begin{figure}[t]
\centering
\includegraphics[width=\textwidth]{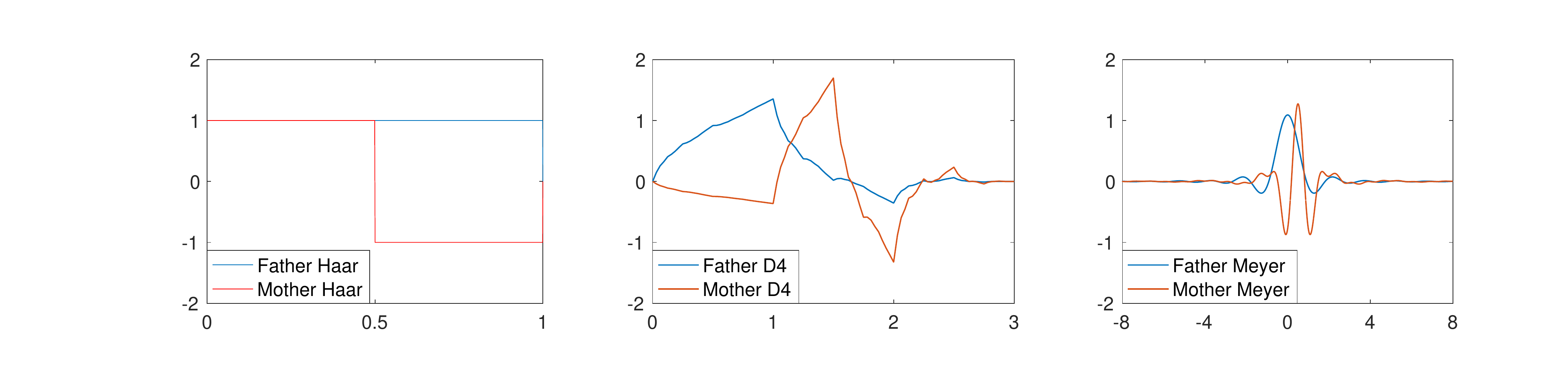}
\caption{Representation of three different wavelets. The Haar wavelet has a compact support and a closed form expression, the Daubechies D4 wavelet has a compact support only and the Meyer wavelet has a closed form expression only. }
\label{fig:wav_bases}
\end{figure}
%%%%%%%%%%%%%%%%%%%%%%%%%%%%%%%%%%%%%%%%%%%%%%%%%%%%%%%%%%%%%%%%%%%%%%%%%
%%%%%%%%%%%%%%%%%%%%%%%%%%%%%%%%%%%%%%%%%%%%%%%%%%%%%%%%%%%%%%%%%%%%%%%%%
%%%%%%%%%%%%%%%%%%%%%%%%%%%%%%%%%%%%%%%%%%%%%%%%%%%%%%%%%%%%%%%%%%%%%%%%%

\subsection{Continuous time wavelet estimator of the intensity }
\label{sec:wavest}

Consider a point process with a piecewise continuous intensity function $\lambda\in L^2(\mathbb{R})$, typically restricted to a finite length observation window $[0,T)$.  We write the following wavelet expansion for this intensity \citep{DeMiranda2011,brillinger1997some}:
\begin{equation}\lambda(t) = \sum\limits_{k \in \mathbb{Z}}\alpha_{j_{0},k}\phi_{j_{0},k}(t)+\sum\limits_{j\geq j_0}\sum\limits_{k \in \mathbb{Z}}\beta_{j,k}\psi_{j,k}(t),\label{wexp}\end{equation}
where $j_{0}\in\mathbb{Z}$ is fixed and called the coarse resolution level, $\alpha_{j_{0},k} = \langle \lambda,\phi_{j_{0},k}\rangle$ and $\beta_{j,k} = \langle \lambda,\psi_{j,k}\rangle$. %(XXX fix this next bit). We also want to use the wavelet analysis of the intensity to form a first-order multiscale analysis of the underlying point process. In Section \ref{sec:test}, this will translate into defining properties in a multiscale fashion and testing them using distributional results for the wavelet expansion \eqref{wexp}. 
We are  required to estimate the coefficients $\alpha_{j_{0},k}$ and $\beta_{j,k}$ which we do so with 
$
\widehat{\alpha}_{j_{0},k} = \int\phi_{j_{0},k}(t)\d N(t) =  \sum_{\tau_{i}\in\mathcal{E}}\phi_{j_{0},k}(\tau_{i})$ and $\widehat{\beta}_{j,k} = \int\psi_{j,k}(t)\d N(t) = \sum_{\tau_{i}\in\mathcal{E}}\psi_{j,k}(\tau_{i}) ,
$ where $ \mathcal{E} = \{ \tau_{i}, 1 \leq i \leq N(T) \}$ are the event times for one realization of a point process $N$ on the time interval $(0,T]$. Hence the general linear estimator of the intensity function based on its wavelet expansion is:
\begin{equation}
\label{fulllinearest}\widehat{\lambda}(t) = \sum\limits_{k \in \mathbb{Z}}^{}\widehat{\alpha}_{j_{0},k}\phi_{j_{0},k}(t)+\sum\limits_{j\geq j_0}\sum\limits_{k \in \mathbb{Z}}\widehat{\beta}_{j,k}\psi_{j,k}(t).
\end{equation}
For a compactly supported wavelet function, Campbell's theorem \citep[Chapter 6]{Daley1998} gives us
\begin{align*}
E\{\widehat{\alpha}_{j_{0},k}\} &= \int\phi_{j_{0},k}(t) E\left\lbrace \d N(t)\right\rbrace =  \int\phi_{j_{0},k}(t)\lambda(t)\d t =\alpha_{j_0,k} \\
E\{\widehat{\beta}_{j,k}\} &= \int\psi_{j,k}(t)E\left\lbrace \d N(t)\right\rbrace = \int\psi_{j,k}(t)\lambda(t)\d t = \beta_{j,k},
\end{align*}
showing the coefficient estimators to be unbiased. This is a linear estimator as it involves no shrinkage of the coefficients. 

For obvious computational reasons, we can not in practice use an infinite wavelet basis to reconstruct the intensity (the intensity may only be fully reconstructed when we know that its decomposition is actually finite). Therefore, we firstly have to choose a maximum resolution level $J$. This maximum level plays a role in the bias-variance tradeoff of the estimator. Low values of $J$ result in a smooth (high bias, low variance) estimator, whereas large values of $J$ result in a noisy (low bias, high variance) estimator. The linear estimator then becomes the estimator of the projection of $\lambda$ onto the space $V_{J} = V_{j_{0}} \oplus  \bigoplus\limits_{j = j_{0}}^{J-1}W_{j}$, and is noted $\widehat{\lambda}^{J}$ from now on. In practice we usually set the coarsest level of resolution $j_{0}$ to 0. Also, with compactly supported wavelets and events restricted to a finite length observation window $[0,T)$, the subset of translation indexes $k \in\mathbb{Z}$ satisfying $\widehat{\beta}_{j,k} \neq 0$ is finite.

 %Since the collection of functions $\{ \phi_{j,k}, k \in \mathbb{Z} \}$ is an orthonormal basis of $V_{j}$ and noting $\widehat{\lambda}^{J}$ as the linear estimator of the projection of $\lambda$ onto $V_{J}$, we can therefore write:
%\begin{equation}
%\label{linearest}
%\widehat{\lambda}^{J}(t) = \sum\limits_{k \in \mathbb{Z}}^{}\widehat{\alpha}_{J,k}\phi_{J,k}(t).
%\end{equation}

A non-linear estimator is obtained by adding a coefficient shrinkage term, determined from a thresholding strategy. The use of shrinkage methods in the classical wavelet regression setting is well studied \cite[e.g.][]{Donoho1995} and is used as a smoothing method to suppress contributing terms from fine scales which typically contain noise. For point process intensity estimation, while we do not have a noise term per se, shrinkage strategies are again desirable for smoothing, with fine scale terms typically having high variance. 

When reconstructing the intensity of a point process, we have two desirable properties for a wavelet function. The first is that it should have a closed-form expression; it will be shown that this is required to compute the estimator of the intensity function. Second, the wavelet should be compactly supported; this is because invariably we can only observe the point process on a finite interval and therefore compactly supported wavelets allow us to only consider a finite set of dyadic translations. In Figure \ref{fig:wav_bases} we show three examples of wavelet families; these are the Haar, Daubechies D4 and Meyer wavelets. Each family exhibits either one or both characteristics.

\subsubsection{Haar estimator}
\begin{figure}[t]
\centering
\includegraphics[width=\textwidth]{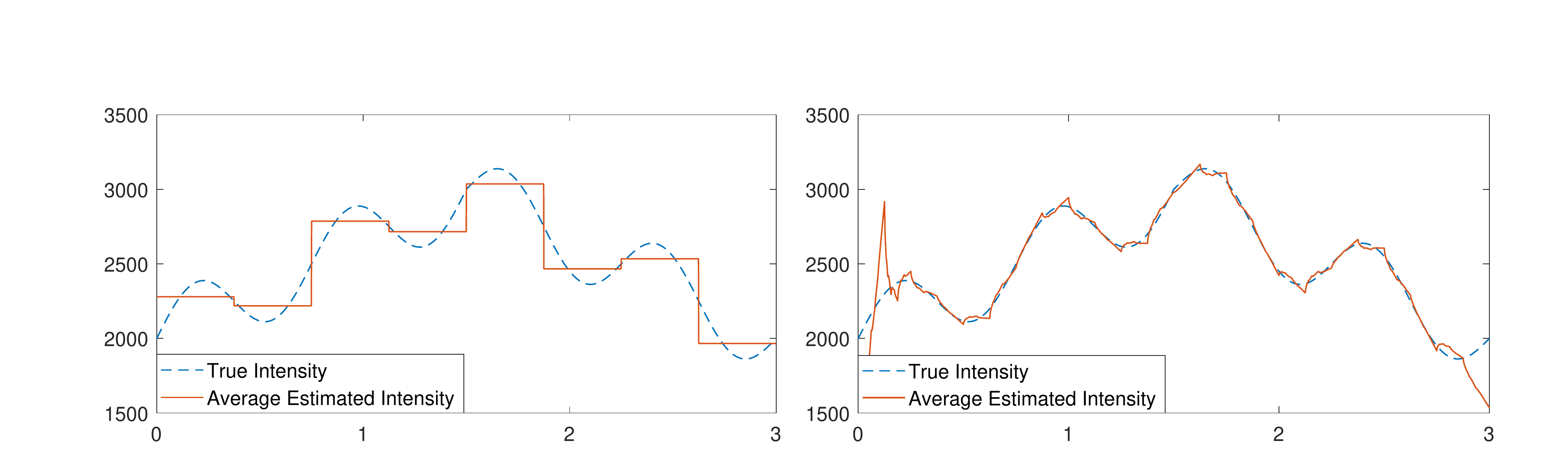}
\caption{Estimation of an example intensity with Haar and D4 wavelets obtained with an average over 1000 realizations of a point process on $[0,3]$. We choose $J = 3$ here. The intensity is the sum of a triangular and a sine function.}
\end{figure}
The Haar mother and father wavelets are defined as \\

\begin{minipage}[b]{0.45\linewidth}
\[\psi(t) =
  \begin{cases}
 1 &\textrm{  if  } 0\leq t < 1/2 \\
 -1  &\textrm{  if  } 1/2\leq t < 1\\
 0 &\textrm{  otherwise  } 
  \end{cases}
\] 
\end{minipage}
and
\begin{minipage}[b]{0.45\linewidth}
\[\phi(t) =
  \begin{cases}
 1 &\textrm{  if  } 0\leq t < 1 \\
 0 &\textrm{  otherwise  } 
  \end{cases}.
\]
\end{minipage}\\\\
These wavelets can be extended to the support $[0,T)$ with an orthonormality preserving rescaling $\psi_T(t) = T^{-1/2}\psi(t/T)$ and dyadic transforms of the type $\psi_{T,j,k}(t) = 2^{j/2}\psi_{T}(2^jt-kT)$. Henceforth, we will drop the subscript $_T$ and assume all wavelets are scaled for the support $[0,T)$. Now consider a point process $N$ on $[0,T)$. By construction, Haar wavelets have disjoint supports across all translations for a fixed scale, which justifies simplification of the indexes. The linear estimator of the intensity function based on its Haar wavelet expansion becomes:
\[\widehat{\lambda}^{J}(t) = \sum\limits_{k =0 }^{2^{j_{0}}-1}\widehat{\alpha}_{j_{0},k}\phi_{j_{0},k}(t)+\sum\limits_{j =j_{0}}^{J-1}\sum\limits_{k = 0}^{2^{j}-1}\widehat{\beta}_{j,k}\psi_{j,k}(t) = \sum\limits_{k =0 }^{2^{J}-1}\widehat{\alpha}_{J,k}\phi_{J,k}(t).\]  
\begin{remark}
\label{remark_haar}
Under the Haar wavelet basis, at scale $J \geq 0$ and a translation $0\leq k \leq 2^J-1$ we have $\alpha_{J,k} = \frac{1}{\sqrt{2}}(\alpha_{J+1,2k} +\alpha_{J+1,2k+1} )$. 
\end{remark}
See proof in Appendix \ref{proof_rem1}. The linear estimator based on the Daubechies D4 wavelets is discussed in Appendix \ref{D2Qest}.

\subsubsection{Distribution of $\widehat\lambda^J$}
\label{sec:dist_lambdaest}
In the case of Haar wavelets, we are able to derive the distribution of the estimator $\widehat\lambda^J$. The approximation space of interest, $V_J$, naturally induces a subdivision $S_{J} = \left\lbrace s^{J}_{k}\right\rbrace_{k=0}^{2^{J}-1}$ of the interval $[0,T)$. The elements of this subdivision, $s^{J}_{k}=[T\frac{k}{2^{J}},T\frac{k+1}{2^{J}})$, are the supports of the Haar wavelets at scale $J$ and form $2^{J}$ disjoint subintervals of  $[0,T)$. The Haar reconstruction of the  intensity $\lambda^J$ and its linear estimator $\widehat\lambda^J$ are piecewise constant functions, with forms $\lambda^J(t) = \sum_{k=0}^{2^J -1}\lambda_k^J\mathbbm{1}_{s_k^J}(t)$ and $\widehat\lambda^J(t) = \sum_{k=0}^{2^J -1}\widehat\lambda_k^J\mathbbm{1}_{s_k^J}(t)$, respectively. Hence we can establish the exact distribution for this estimator under a Poisson process model.

\begin{proposition}
\label{prop}
Under the Haar wavelet basis and for an inhomogeneous Poisson process $N$ of intensity $\lambda$ on $[0,T)$, $\widehat\lambda_0^J,...,\widehat\lambda_{2^J -1}^J$ are independent random variables distributed as $$ \widehat{\lambda}^{J}_{k} \sim \frac{2^{J}}{T}\text{\rm Pois}(\mu^{J}_{k}), \quad 0\leq k \leq 2^{J}-1,$$ where $\mu^{J}_{k} = \int\limits_{s^{J}_{k}}\lambda(t)\d t$.
\end{proposition}
The proof can be found in Appendix \ref{proof_prop}. The result can also naturally be extended to any other point process with a square integrable intensity function for which the distribution of the event counts in any time interval is known (e.g. a binomial point process). It follows that $E\{\widehat{\lambda}^{J}_{k}\} = \lambda^{J}_{k} = \frac{2^{J}}{T}\mu^{J}_{k}$, for all $0\leq k \leq 2^{J}-1$. We will now use Proposition \ref{prop} to develop likelihood ratio tests for two newly defined multiscale properties of a Poisson process. 
%%%%%%%%%%%%%%%%%%%%%%%%%%%%%%%%%%%%%%%%%%%%%%%%%%%%%%%%%%%%%%%%%%%%%%%%%
%%%%%%%%%%%%%%%%%%%%%%%%%%%%%%%%%%%%%%%%%%%%%%%%%%%%%%%%%%%%%%%%%%%%%%%%%
%%%%%%%%%%%%%%%%%%%%%%%%%%%%%%%%%%%%%%%%%%%%%%%%%%%%%%%%%%%%%%%%%%%%%%%%%

\section{A New Testing Protocol for Multiscale Properties of Poisson Processes}
\label{sec:test}
\setcounter{figure}{0}  

In this section, we will develop the theoretical framework for a wavelet-based multiresolution analysis of a point process. Considering the first order properties of a point process to be due to activity on different scales, under the Haar basis we define different levels of homogeneity under a multiresolution framework. We call this $J$-th level homogeneity, and provide a likelihood ratio test for it for the class of Poisson processes. 

Under a compactly supported wavelet family, we then consider a more general setting to describe any activity of the intensity function at a particular scale, which we term $L$-th level innovation. We provide a likelihood ratio test for this property for the class of Poisson processes under the Haar basis. In Section \ref{sec:thresh}, we will demonstrate how this test can be used as a method of thresholding coefficients in our wavelet estimator of the intensity function. In this section, it will be always assumed that the intensity $\lambda$ is piecewise continuous and $\lambda \in L^2(\mathbb{R})$. 

\subsection{Global behaviour: $J$-th level homogeneity}

We use the Haar wavelet basis (rescaled if T is different than 1), because of its intuitive interpretation, its simplicity to implement and its amenability to statistical analysis. We consider the projection of the intensity on the Haar approximation space $V_{J} = V_{j_{0}} \oplus \bigoplus\limits_{j = j_{0}}^{J-1}W_{j}$. With Haar wavelets, the  reconstruction of the intensity at scale $J$ is a piecewise constant function, and hence we can define a wavelet reconstruction vector $(\lambda_{0}^{J},\lambda_{1}^J,\cdots,\lambda_{2^{J}-1}^{J})^T$ where $\lambda_{k}^{J}$ is the value of $\lambda^J$ on the subinterval $s^{J}_{k}\in S_J$, , $k=0,...,2^J -1$. We use this formulation to define a property we call $J$-th level homogeneity.
\begin{definition}
\label{jhomo} A point process $N$ on $[0,T)$ with intensity $\lambda$ is considered \textbf{level $J$ homogeneous} if the reconstruction of the intensity at resolution $J$ with Haar wavelets, or its projection on $V_{J}$, is constant on $[0,T)$. That is, $\lambda_{0}^{J}=\lambda_1^J=...=\lambda_{2^{J}-1}^J$.
\end{definition}
$J$th-level homogeneity was introduced in \cite{Taleb2016} in terms of the projection of the intensity on $V_{J+1}$. We propose that it is instead more convenient to base it on $V_{J}$, i.e. every point process is level $0$ homogeneous as the projected intensity $\lambda^0_0$ on $V_{0}$ is always a constant on $[0,T)$. The concept of $J$-th level homogeneity goes side by side with the idea of a multiresolution analysis of the intensity function, providing a natural way of studying on what scales the intensity function appears constant and hence the point process homogeneous, and on what scales the intensity function exhibits variability. If we define $H_{J}$ as the class of level $J$ homogeneous point processes, we have $H_{J} \supset H_{J+1}$. Indeed we know from Remark \ref{remark_haar} that $\alpha_{J,k} = \frac{1}{\sqrt{2}}(\alpha_{J+1,2k} +\alpha_{J+1,2k+1} )$ for Haar wavelets, and therefore $\lambda_{0}^{J} = \lambda_{1}^{J} = \cdots = \lambda_{2^J -1}^J$ if $\alpha_{J+1,0} = \alpha_{J+1,1} = \cdots = \alpha_{J+1,2^{J+1}-1}$. 
\begin{proposition}
\label{homo_eq}
Let $N$ be a point process with intensity $\lambda$. Then $\lambda$ is constant almost everywhere on $[0,T)$ (i.e. $\lambda(t) = \lambda_0^0  = \frac{1}{T}\int\limits_0^T \lambda(t)\dif t$ almost everywhere) if and only if $N \in H_J $ for all $J \geq 0$. 
\end{proposition}
See Appendix \ref{proof_homo_eq} for the proof. To avoid any confusion, we say that a point process with intensity $\lambda$ is strictly homogeneous on $[0,T)$ when $\lambda(t) = \lambda_0^0$ for all $t \in [0,T)$. Proposition \ref{homo_eq} illustrates how strict homogeneity can be loosely interpreted as the limit extension of $J$th-level homogeneity. Furthermore, Definition \ref{jhomo} naturally leads us to define $J$th-level inhomogeneity. 
\begin{definition}\label{jinhomo} A point process  $N$ on $[0,T)$ with intensity $\lambda$  is considered \textbf{level $J$ inhomogeneous} if it is level $J-1$ homogeneous and not level $J$ homogeneous.
\end{definition}
We immediately remark that a level $J$ inhomogeneous point process cannot be level $j$ homogeneous for all $j \geq J$. $J$-th level homogeneity and inhomogeneity together describe the global behavior of a point process when viewed at a particular scale.

\subsection{Testing $J$-th level homogeneity}

As the scope of this work is to analyse point processes in a multiscale fashion, we are not interested in testing the strict homogeneity of a Poisson process, which is the limit case for Definition \ref{jhomo} and has been thoroughly addressed in previous studies \citep[e.g.][]{Bain1985,Ng1999}. We are instead aiming to statistically determine the resolution level where inhomogeneous behaviour appears. Recall that the choice of Haar wavelets implies that the wavelet reconstruction $\lambda^J$ of the intensity $\lambda$, as well as the intensity estimator $\widehat{\lambda}^{J}$, are  piecewise constant functions on the dyadic partition $S_J$. Although a piecewise analysis has also been carried out in \cite{Fierro2011} as a basis for a similar LRT, the wavelet approach presented here gives a natural, multiresolution scheme for defining the subdivision of the process. We begin by considering the LRT for equal means of scaled Poisson distributions, the results of which we can then utilize to test $J$-th level homogeneity of Poisson processes. This provides a comprehensive and rigorous treatment of the ideas first proposed in \cite{Taleb2016}.

\subsubsection{LRT for equal means of scaled Poisson distributions}
\label{sec:lrt_scaled}
Let $\mathbb{X} = \left\{\mathbf{X}_{m}\right\}_{m = 1}^{M}$ be a set of iid scaled Poisson random vectors, each with independent components of form $\mathbf{X}_{m} = \left(X_{m,i}\right)_{i = 1}^{P}$, $X_{m,i} \sim \delta \rm{Pois}(\mu_i)$. The scale parameter $\delta>0$ is known and fixed so  $\mathbf{X}_{m}$ is parametrized by the vector $\left(\mu_{i}\right)_{i = 1}^{P}$. We consider testing the null hypothesis $H : \mathit{\mu_{1}=\cdots=\mu_{P}=\mu_c}$ against the alternative hypothesis $K$ that states $H$ is not true. The LRT statistic is defined as
\begin{align}
&r = \quad  \frac{\quad \mysup{} {\mu_{c} > 0}\ \L(\mathbb{X};\mu_{c},...,\mu_c) \quad }{\quad \mysup{}{\left\{\mu_{i}\right\}_{i = 1}^{P}, \sum \mu_i > 0}\ \L(\mathbb{X};\mu_1,...,\mu_P) \quad },
\label{lrs_homo}
\end{align}
where $\L(\mathbb{X};\mu_1,...,\mu_P)$ is the likelihood of the data $\mathbb{X}$ given parameter vector $\left(\mu_{i}\right)_{i = 1}^{P}$.
\begin{proposition}
\label{prop_Rstat}
Let $R=-2\log\left(r\right)$, with $r$ being the likelihood ratio statistic defined in (\ref{lrs_homo}). Then we have $$R=2M\sum\limits_{i = 1}^{P}\bar{\mu}_{i}\log\left(\frac{\bar{\mu}_{i}}{\bar{\mu}_{c}}\right),$$ where $\bar{\mu}_{c}  = \frac{1}{\delta MP}\sum\limits_{i = 1}^{P}\sum\limits_{m= 1}^{M}X_{m,i}$ is the maximum likelihood estimator (MLE) for $\mu_c$, the constant mean under the null hypothesis $H$, and $\bar{\mu}_{i} = \frac{1}{\delta M}\sum\limits_{m= 1}^{M}X_{m,i}$ is the MLE for $\mu_{i}$ ($i=1,...,P$), under the alternative hypothesis $K$.
\end{proposition}
See Appendix \ref{proof_Rstat} for the proof. If there exists at least one index $i$ such that $\bar{\mu}_{i} = 0$, we use the convention $0 \log(0)= 0$. Further discussion on the absence of points within intervals can be found in Appendix \ref{nopoints}. Now let $d_{H}$ be the number of free parameters under the null hypothesis $H$ and let $d_{K}$ be the number of free parameters under the alternative hypothesis $K$, then under the null hypothesis and  regularity conditions on the likelihood functions that are met here, $R\rightarrow \chi^2_{d_K - d_H}$ as sample size $M\rightarrow \infty$ \citep[see][]{Wilks1938,Van2000}. In this setting, $d_K = P$ and $d_H = 1$. In practice, the $M = 1$ case is frequently encountered, and therefore we establish a more general and applicable result for the asymptotic distribution of $R$. 
\begin{theorem}
\label{chi2}
Let $\bX_1,..,\bX_M$ ($M\geq 1$) be independent and identically distributed $P$ dimensional random vectors where each $\bX_m = \left(X_{m,1},...,X_{m,P}\right)^T$ is constructed from independent components $X_{m,i}\sim \delta\ \rm{Pois}(\mu_i)$. Let $R = -2\log(r)$ where $r$ is the likelihood ratio statistic defined in (\ref{lrs_homo}). Then the distribution of statistic $R$ is invariant to simultaneous changes in parameters $M$ and $\mu_{i}$ provided that all products $\mu_{i}M$ , $1 \leq i \leq P$, remain constant. Furthermore, if $\mu_1 = ... =\mu_P = \mu_c$, then $R \overset{d}{\rightarrow} \chi^{2}_{P - 1}$ as $\mu_{c}M \rightarrow \infty$. 
\end{theorem}
See Appendix \ref{proof_chi2} for the proof\footnote{It has been shown in \cite{Feng2012} that the classic asymptotic distributional result for the test statistic $R$ does not hold if we are restricting ourselves to the $M = 1$ case and low values of $\mu_c$ ($\mu_c \leq 10$ in their study). This refutes the opposite claim in \cite{Brown2002}, which possibly resulted from a confusion between the number of parameters $P$ and the number $M$ of independent realizations of the Poisson vector.}. It will now be shown that this result illustrates the practical advantage of Haar wavelets as it ensures that only one realization of the process is enough to conduct a LRT for $J$-th level homogeneity. 

\subsubsection{LRT for $J$-th level homogeneity of a Poisson process}
Now let $\{N_m,m=1,...,M\}$ be a collection of $M\geq 1$ independent realizations of the same Poisson process $N$. Let $\Lambda = \left\{\bLambda_{m}\right\}_{m = 1}^{M}$ be the set of $M$ independent random vectors where $\bLambda_{m} = \left(\widehat{\lambda}_{m,k}^{J}\right)_{k = 0}^{2^{J}-1}$ is the vector of all subinterval estimates of the intensity from $N_{m}$. From Proposition \ref{prop}, $\bLambda_{m}$ is a vector of independent scaled Poisson random variables and is therefore parametrized by the vector $\left(\lambda_{k}^{J}\right)_{k = 0}^{2^{J}-1}$. We look to test the null hypothesis $H$ which states $N$ is level $J$ homogeneous, i.e. $\lambda_0^J=\cdots=\lambda_{2^J-1}=\lambda_c^J$ for some $\lambda_c^J>0$, against the alternative hypothesis $K$ which states $H$ is not true. The LRT statistic in this case is given as:
\begin{equation*}r^J = \frac{\mysup{}{\lambda^J_{c} > 0}\ \L(\Lambda;\lambda^J_{c},...,\lambda^J_c)}{\mysup{}{\left\{\lambda_{k}^J\right\}_{k = 0}^{2^J-1}, \sum \lambda_k^J > 0}\ \L(\Lambda;\lambda_0^J,...,\lambda_{2^{J}-1}^J)} ,\end{equation*}
where $\L(\Lambda;\lambda_0^J,...,\lambda_{2^{J}-1}^J)$ is the likelihood of the data $\Lambda$ given parameter vector $\left(\lambda_{k}^{J}\right)_{k = 0}^{2^{J}-1}$. Now using Proposition \ref{prop_Rstat} we can write
\[R^J= -2\log(r^J) = 2\frac{M}{\delta^{J}}\sum\limits_{k = 0}^{2^{J}-1}\bar{\lambda}_{k}^{J}\log\left(\frac{\bar{\lambda}_{k}^{J}}{\bar{\lambda}_{c}^J}\right),\]
where $\delta^{J} = 2^{J}/T$, statistic $\bar{\lambda}_{c}^J$ is the maximum likelihood estimator (MLE) of $\lambda^J_c$ and $\bar{\lambda}_{k}^{J}$ is the MLE for $\lambda_{k}^{J}$ ($k=0,...,2^{J}-1$), under the alternative hypothesis $K$. In this particular setting we have $d_K = 2^{J}$ and $d_H = 1$, giving $R$ as asymptotically $\chi^{2}$ distributed with $2^{J} - 1$ degrees of freedom under the conditions of Theorem \ref{chi2}. We reject $J$-th level homogeneity at significance level $\alpha$ if $R>c_\alpha$ where $c_{\alpha}$, the critical value, is the upper $100(1-\alpha)\%$ point of the $\chi^{2}_{2^{J} - 1}$ distribution.

\subsubsection{Simulation study}
\label{sec:errors}

\begin{figure}[t]
\centering
\includegraphics[width=\textwidth]{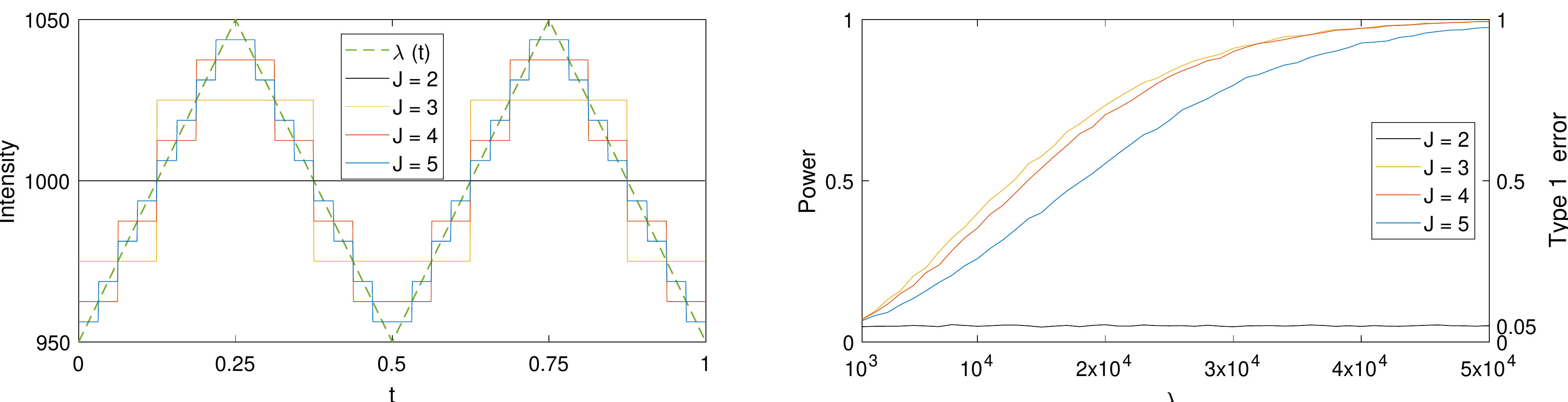}
\caption{\textit{Left}: Haar wavelet reconstruction of a piecewise triangular intensity with $V = 1$, $\xi = 1/10$ and $T = 1$ at resolutions $J \in \{2,3,4,5\}$. \textit{Right}: Empirical type 1 error ($J=2$) and power ($J\in \left\{3,4,5\right\}$) for this piecewise triangular intensity as a function of $\lambda_0$.} 
\label{fig:lrt_homo}
\end{figure}

Here, we demonstrate the LRT for $J$-th level homogeneity through simulations. We consider a class of inhomogeneous Poisson processes on a time interval $[0,T)$. These processes share a similar piecewise triangular intensity represented in Figure \ref{fig:lrt_homo} and are defined as following:
\begin{equation*}
\lambda(t) = \lambda_0\left(\frac{2-\xi}{2} - s(t)(i(t)\bmod 2)\xi\right)  + s(t)a\left(t-\frac{i(t)}{2^{V+1}}T\right),
\end{equation*}
where $s(t) =  1 - 2(i(t)\bmod 2)$ and $a = \frac{2^{V+1}}{T}\xi\lambda_0$,
and $i(t) \in \left\{0, \dots, 2^{V+1}-1\right\}$ is the index of the subinterval $s_{i(t)}^{V+1} = [\frac{i(t)}{2^{V+1}}T,\frac{i(t)+1}{2^{V+1}}T)$ in which $t$ belongs. Parameter $a$ is the absolute value of the gradient and $2^{V}$ is the number of ``triangles''. The intensity takes values between $\lambda_{0}\frac{2-\xi}{2}$ and $\lambda_{0}\frac{2+\xi}{2}$ and its mean value $\lambda_{0}^0$ is the parameter $\lambda_{0} > 0$. By construction, the quantity $\mu_0^0 = \int\limits_{0}^{T}\lambda(t)dt = T\lambda_0$ does not depend on $V$, the process is level $V+1$ homogeneous and level $V+2$ inhomogeneous. We set the significance level of our test at $\alpha=0.05$, with $M = 1$, i.e. we observe just a single realization. The empirical type 1 error and power of the LRT (over 10000 simulations) at different values of $J$ are shown in Figure \ref{fig:lrt_homo} as a function of $\lambda_0$, with $\lambda_0 \in [1000,50000]$.
In the example represented in Figure \ref{fig:lrt_homo} where the process is level $2$ homogeneous, the empirical type 1 error lies close to the $5\%$ level as expected. When $J \geq 3$ and $J$-th level homogeneity no longer holds, the empirical power converges to 1 when $\lambda_{0} \rightarrow \infty$. This behavior is expected as well. Indeed, this intensity model is proportional to $\lambda_0$ and therefore its Haar reconstruction at any scale $J$ satisfies $\lambda_{k}^{J} \propto \lambda_0$ as well as $\mu_{k}^{J} \propto \lambda_0$. Since statistic $R^J$ tends to infinity as $M$ increases towards infinity when $J \geq 3$ and a fixed $\lambda_0$, then the power of the LRT converges to 1. Hence the observed convergence of the empirical power to 1 when $M$ is fixed and $\lambda_0$ increases towards infinity as ensured by Theorem \ref{chi2}. Similarly, the value of parameter $\xi$ influences the speed of this convergence. Moreover, we note the power decreases as we increase $J$ because the mass of the null distribution $\chi^2_{2^{J} -1}$ is displaced to the right as $J$ increases, making it harder for the test to distinguish between the two hypotheses.

\subsection{Local behaviour: $L$-th level innovation}

In Section \ref{wavelet}, we presented the decomposition $L^{2} (\mathbb{R}) = \overline{V_{j_{0}} \oplus \bigoplus\limits_{j = j_{0}}^{\infty}W_{j}}$ where $W_{j}$ is the orthogonal complement of $V_{j}$ in $V_{j+1}$ and often called the detail or innovation space. With $J$-th level homogeneity we focused on the behavior displayed on any space $V_{j}$, which brings together contributions from several resolutions. Projecting $\lambda$ on $W_j$ for increasing $j\geq j_0$, we explore the intensity function in progressively finer resolutions. To characterize this, we introduce the concept of $L$-th level innovation. We consider a wavelet family $(\phi,\psi)$ with compact support, and a point process $N$  on $[0,T)$. For a particular scale $L$, we note $\widetilde{W_{L}}$ the subspace of the detail space $W_{L}$ generated by dyadic transformations $\psi_{L,k}$ of the mother wavelet whose support is included in $[0,T)$. For example, for Haar wavelets, $\widetilde{W_{L}}= {\rm span}\{\psi_{L,k};k=0,...,2^L -1\}$.
 \begin{definition}
  \label{innov}
Let $N$ be a point process with intensity $\lambda$ and  let $(\phi,\psi)$ be a compactly supported wavelet family. We then say that $N$ possesses a \textbf{level L innovation under $\boldsymbol(\boldsymbol \phi,\boldsymbol\psi\boldsymbol)$} if and only if there exist $k\in\mathbb{Z}$ such that $\psi_{L,k} \in \widetilde{W_{L}}$ and $\beta_{L,k} = \langle\lambda,\psi_{L,k}\rangle \neq 0.$
\end{definition}
 The justification behind only considering $\widetilde{W_L}$ is motivated by the analysis of homogeneous Poisson processes. Since a homogeneous Poisson process is level $J$ homogeneous for all $J \geq 0$, we desire that it similarly displays no $L$-th level innovation irrespective of $L \geq 0$ and the wavelet family used. With a constant intensity on observation window $[0,T)$, wavelets with non compact support will always produce an infinite number of non-zero wavelet coefficients and unbiasedness of their estimators is not guaranteed. Furthermore, compactly supported wavelets whose support is only partially contained within $[0,T)$ will also admit non-zero wavelet coefficients. %Then, we also want to avoid the consideration of non-zero wavelet coefficients that are solely due to boundary effects when the intensity has a compact support. Indeed, a constant intensity on $[0,T)$ yields non-zero mother wavelet coefficients for a Daubechies wavelet whose support is partially included in on $[0,T)$. From the cancellation property of wavelets, the absence of $L$th-level innovation defined from the space $\widetilde{W_{L}}$ and not $W_L$ is guaranteed for any compactly supported wavelet in the case of a homogeneous Poisson process. 
$L$-th level innovation is dependent on the wavelet family used to reconstruct the intensity. In the case of Haar wavelets, we will show in Section \ref{testinginnovation} it has an intuitive interpretation as the absence of any change in the integrated intensity between the left and right hand sides of the Haar wavelet. Admittedly, such an interpretation becomes less intuitive with alternative wavelets. We further comment that although defined according to a specific scale, $L$-th level innovation also has an inherent temporal component. The translation index of non-zero coefficients given by wavelets in $\widetilde{W_{L}}$ indicates the time localization of the corresponding innovation.
\begin{remark}
\label{rem2}
For the Haar wavelet, there is the following equivalence:
\begin{itemize}
\item  A point process $N$ is level $J$ homogeneous and possesses a level $J$ innovation.
\item  A point process $N$ is level $J+1$ inhomogeneous.
\end{itemize}  
\end{remark}
This equivalence is immediate from applying Definitions \ref{jinhomo} and \ref{innov} to the identity $V_{J+1} = V_{J} \oplus W_{J}$. 

\subsection{Testing $L^{th}$-level innovation}
\label{testinginnovation}
We are now interested in testing for $L$-th level innovation based on Definition \ref{innov} using the null hypothesis $H$: ``\emph{A point process $N$ possesses no $L$-th level innovation under a wavelet family $(\phi,\psi)$}''. To do so, we consider the vector of empirical wavelet coefficients corresponding to the wavelet basis for $\widetilde{W_L}$, which under the null hypothesis will be zero mean. %Since an innovation in the sense of one wavelet basis does not carry the same meaning as an innovation in the sense of another one, detecting one and not the other should not be a basis for performance ranking. Instead, different hypothesis tests let us explore different wavelet-dependent interpretations of the multiscale behaviour of a point process. 
As for $J$-th level homogeneity, we define a likelihood ratio test for $L^{th}$-level innovation under the Poisson process model and Haar wavelets. This test will again be a special case of a more general setting for multivariate Poisson random variables. 

If a point process is level $J$ inhomogeneous, then such a test should take place for any given scale $L > J$ (as by Remark \ref{rem2} we know there must be innovation at level $J$). Consider a subdivision $S_{L+1}$ of $[0,T)$ defined as in Section \ref{sec:dist_lambdaest}. Let $\{N_m\}_{m = 1}^{M}$ be a collection of $M$ independent realizations of the same Poisson process $N$ on $[0,T)$ with intensity function $\lambda$, and let $\mathbb{X}_N = \left\{\bX_{m}\right\}_{m = 1}^{M}$ be a collection of $M$ independent random vectors $\bX_{m} = \left(X_{m,i}\right)_{i = 0}^{2^{L+1}-1}$, where $X_{m,i} = N_m(s_i^{L+1})$ is the event count for process $N_m$ in $s_i^{L+1} \in S_{L+1}$. With $\widehat{\beta}_{L,k} = \sum_{\tau_{i}\in\mathcal{E}}\psi_{L,k}(\tau_{i})$, for the Haar wavelets $\widehat{\beta}_{L,k} = \frac{2^{L/2}}{\sqrt{T}}(X_{m,2k}-X_{m,2k+1}), 0 \leq k \leq 2^L-1$. Each count $X_{m,i}$ is distributed as $\text{\rm Pois}(\mu_{i})$ where $\mu_{i} = \int_{s_i^{L+1}}\lambda(t)\d t$. Therefore, the estimators $\widehat{\beta}_{L,k}$, $k=0,...,2^{L}-1$ are independent realizations of a scaled Skellam distribution (or Poisson difference distribution), each with parameters $\mu_{2k}$ and $\mu_{2k+1}$. Since $\widehat{\beta}_{L,k}$ has mean $ \frac{2^{L/2}}{\sqrt{T}}(\mu_{2k}-\mu_{2k+1})$, Definition \ref{innov} is then equivalent to the following property: ``\emph{There exist $k\in 0,...,2^L -1$ such that $\widehat{\beta}_{L,k}$ is Skellam distributed with parameters $\mu_{2k}\neq\mu_{2k+1}$}''.  We can therefore build a likelihood ratio test for testing the null hypothesis $H$: ``\emph{$\mu_{2k}= \mu_{2k+1}$ for all $k=0,...,2^L -1$}''. 

Since there does not exist an explicit expression for the MLE of the parameter $\theta_k = \mu_{2k} - \mu_{2k+1}$ given Skellam distributed random variables (instead having to be numerically approximated \citep{alzaid2010poisson}), it is more appealing to design a likelihood ratio test based on the event counts themselves. This leads us to first consider a LRT for the general setting of testing pairwise equality of means of Poisson distributions, which will then be used for the specific setting of testing $L$-th level innovation.

\subsubsection{LRT for pairwise equality of Poisson means}
\label{sec:lrt_pairs}
We define here a LRT for the pairwise equality of the means of a multivariate Poisson distribution. Let $\mathbb{X} = \left\{\mathbf{X}_{m}\right\}_{m = 1}^{M}$ be a set of iid Poisson random vectors, each with independent components of form $\mathbf{X}_{m} = \left(X_{m,i}\right)_{i = 1}^{2P}$, $X_{m,i} \sim \rm{Pois}(\mu_i)$.  We consider testing the null hypothesis $H : \mathit{\mu_{2i-1}=\mu_{2i} = \mu_i^{pair},\ 1 \leq i \leq P}$, against the alternative hypothesis $K$ that states $H$ is not true. The LRT statistic is defined as
\begin{align}
&r =  \frac{\hspace{1.5cm} \mysup{} {\left\{\mu_{i}^{pair}\right\}_{i = 1}^{P},\ \sum \mu_i^{pair} > 0}\ \L(\mathbb{X};\mu^{pair}_{1},\mu^{pair}_{1},...,\mu^{pair}_{P},\mu^{pair}_{P}) \quad }{\hspace{1.5cm} \mysup{}{\left\{\mu_{i}\right\}_{i = 1}^{2P},\ \sum \mu_i > 0}\ \L(\mathbb{X};\mu_1,\mu_2,...,\mu_{2P-1},\mu_{2P}) \quad },
\label{lrs_innov}
\end{align}
where $\L(\mathbb{X};\mu_1,...,\mu_{2P})$ is the likelihood of the data $\mathbb{X}$ given parameter vector $\left(\mu_{i}\right)_{i = 1}^{2P}$.

\begin{proposition}
\label{prop_Rstat_innov}
Let $R=-2\log\left(r\right)$, with $r$ being the likelihood ratio statistic defined in (\ref{lrs_innov}). Then $$R =2M\left[\sum\limits_{i = 1}^{P}\bar{\mu}_{2i-1}\log\left(\frac{\bar{\mu}_{2i-1}}{\bar{\mu}_{i}^{pair}}\right)+\sum\limits_{i = 1}^{P}\bar{\mu}_{2i}\log\left(\frac{\bar{\mu}_{2i}}{\bar{\mu}_{i}^{pair}}\right)\right],$$ where $\bar{\mu}_{i} = \frac{1}{M}\sum_{m=1}^MX_{m,i}$ and $\bar{\mu}_{i}^{pair} = \frac{1}{M}\sum_{m=1}^M\widehat{\mu}_{m,i}^{pair}$ where $\widehat{\mu}_{m,i}^{pair} = \frac{1}{2}(X_{m,2i-1}+X_{m,2i})$. 
Statistic $\bar{\mu}_{i}^{pair}$ is the maximum likelihood estimator (MLE) of $\mu_{i}^{pair}$ ($i=1,...,P$) under the null hypothesis $H$ and $\bar{\mu}_{i}$ is the MLE for $\mu_{i}$ ($i=1,...,2P$) under the alternative hypothesis $K$.
\end{proposition}
The proof can be found in Appendix \ref{proof_Rstat_innov}. From Wilks' Theorem \citep{Wilks1938}, we immediately have that under the null hypothesis $R$ is asymptotically $\chi^{2}$ distributed with $d_{K}-d_{H} = P$ degrees of freedom for a large sample size $M$ (under the usual regularity assumptions). However, this result is not guaranteed when the true parameter vector lies on the boundary of the parameter space. This was not the case for the test in Section \ref{sec:lrt_scaled} since we must have $\mu_c > 0$, although it happens in this model when $\mu^{pair}_{i}=0$. Further discussion on this particular case can be found in Appendix \ref{ap:boundary}. We now assume that  $\mu^{pair}_{i} \neq 0$ for all $1 \leq i \leq P$. Similarly to Theorem \ref{chi2}, we can state an extension of Wilks' theorem for this LRT.
\begin{theorem}
\label{innov_chi2}
Let $\bX_1,..,\bX_M$ ($M\geq 1$) be independent and identically distributed $P$ dimensional random vectors where each $\bX_m = \left(X_{m,1},...,X_{m,2P}\right)^T$ is constructed from independent components $X_{m,i}\sim \rm{Pois}(\mu_i)$. Let $R=-2\log\left(r\right)$ where $r$ is the likelihood ratio statistic defined in (\ref{lrs_innov}). Then the distribution of statistic $R$ is invariant to simultaneous changes in parameters $M$ and $\mu_{i}$ provided all products $\mu_{i}M$ , $1 \leq i \leq 2P$ remain constant. Furthermore, if $\mu_{2i-1}=\mu_{2i} = \mu_i^{pair} \textrm{ and } \mu^{pair}_{i} \neq 0,\ 1 \leq i \leq P$, then $R \overset{d}{\rightarrow} \chi^{2}_{P}$ as $\mu_i^{pair}M \rightarrow \infty,\ 1 \leq i \leq P$. 
\end{theorem}
The proof of Theorem \ref{innov_chi2} follows an analogous argument to that of Theorem \ref{chi2} (see Appendix \ref{proof_chi2_innov}). We again demonstrate that in the asymptotic analysis of the distribution of $R$, $M$ and the mean intensity are indistinguishable from their product and thus the results are applicable for only one realization of the random vector $\bX$.

\subsubsection{LRT for $L$-th level innovation}
\label{sec:lrtinnov}
We can now apply the test developed in Section \ref{sec:lrt_pairs} to the task of testing $L$-th level innovation. The LRT statistic for testing the null hypothesis $H$: ``\emph{$\mu_{2k}= \mu_{2k+1}$ for all $k=0,...,2^L -1$}'' is
$$r^L = \frac{\hspace{1.5cm} \mysup{} {\left\{\mu_{k}^{pair}\right\}_{k = 0}^{2^L-1},\ \sum \mu_k^{pair} > 0}\ \L(\mathbb{X};\mu^{pair}_{0},\mu^{pair}_{0},...,\mu^{pair}_{2^L-1},\mu^{pair}_{2^L-1}) \quad }{\hspace{1.5cm} \mysup{}{\left\{\mu_{k}\right\}_{k = 0}^{2^{L+1}-1},\ \sum \mu_k > 0}\ \L(\mathbb{X};\mu_0,...,\mu_{2^{L+1}-1}) \quad }.$$
From Proposition \ref{prop_Rstat_innov} we have:
$$R^L = -2\log(r^L) = 2M\left[\sum\limits_{k = 0}^{2^{L}-1}\bar{\mu}_{2k}\log\left(\frac{\bar{\mu}_{2k}}{\bar{\mu}_{k}^{pair}}\right)+\sum\limits_{k = 0}^{2^{L}-1}\bar{\mu}_{2k+1}\log\left(\frac{\bar{\mu}_{2k+1}}{\bar{\mu}_{k}^{pair}}\right)\right].$$
Again, we refer to Appendix \ref{ap:boundary} in the situation where one or several parameters $\mu_k^{pair}$ are equal to zero. In all other cases, we have $d_K = 2^{L+1}$ and $d_H = 2^L$, giving $R$ as asymptotically $\chi^{2}$ distributed with $2^{L} $ degrees of freedom under the conditions of Theorem \ref{innov_chi2}. We reject the absence of a level $L$ innovation at significance level $\alpha$ if $R>c_\alpha$ where $c_{\alpha}$, the critical value, is the upper $100(1-\alpha)\%$ point of the $\chi^{2}_{2^{L}}$ distribution.

\subsubsection{Simulation study}
\label{sec:sim}
\begin{figure}[t]
\centering
\includegraphics[width=\textwidth]{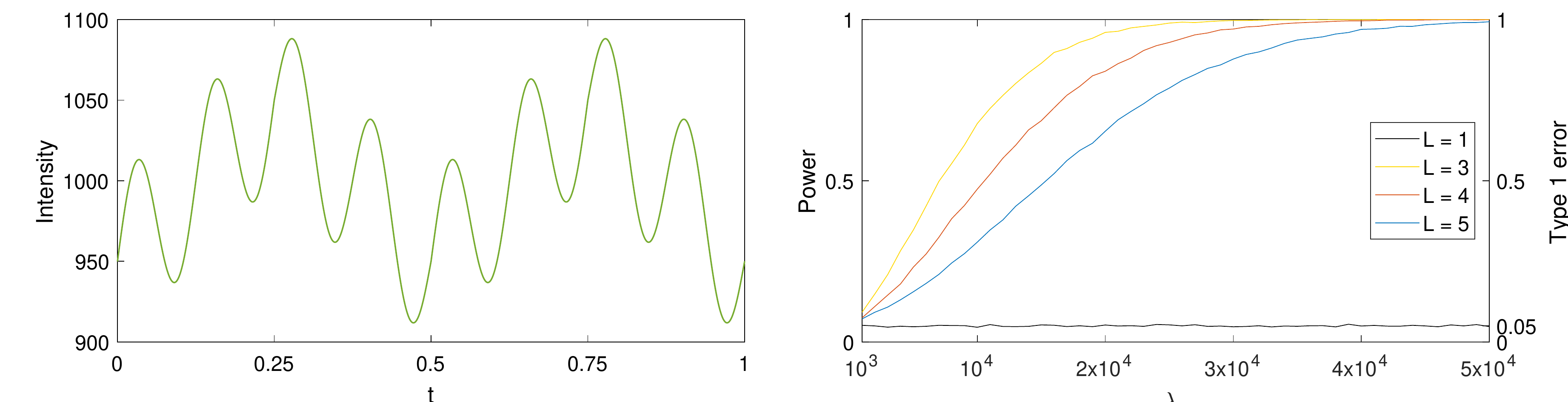}
\caption{\textit{Left:} Triangular rate on $[0,1]$ with mean $\lambda_{0} = 1000$, $V = 1$, $\chi = 0.1$ and an additive sine perturbation with $\nu = 3$ and magnitude $A = 0.05$ . \textit{Right:} Empirical type 1 error ($L = 1$) and power plots ($L\in \left\{3,4,5\right\}$) as a function of $\lambda_0$ with $T = 1$, $V = 1$, $\chi = 0.1$ and $A=0.05$. See text in Section \ref{sec:sim} for further details.}
\label{fig:lrt_innov}
\end{figure}

Let us now consider the triangular intensity model from Section \ref{sec:errors} where we now introduce an additive perturbation in the form of a sine function with period $T/2^\nu, \nu \geq V + 3,$ and magnitude $A\lambda_{0}$. Again, $T$ is the length of the process and $\lambda_{0}$ is the mean value of the rate. Therefore this intensity model has expression
 \begin{align*}
\lambda_{\textrm{sine}}(t) &= \lambda_0\left(\frac{2-\xi}{2} - s(t)(i(t)\bmod 2)\xi\right)  + s(t)a\left(t-\frac{i(t)}{2^{V+1}}T\right) + A\lambda_0\sin\left(\frac{2^{\nu+1}\pi}{T}t\right).
\end{align*}
Similarly to the previous model, the quantity $\mu_0^0 = \int\limits_{0}^{T}\lambda_{\textrm{sine}}(t)dt = T\lambda_0$ does not depend on $V$, the process is level $V+1$ homogeneous and level $V+2$ inhomogeneous. The sinusoidal term does not influence the values of the wavelets coefficients up to resolution $\nu$. Hence a Poisson process $N$ whose intensity is $\lambda_{\textrm{sine}}$ possesses no innovations from levels $0$ to $V$, $V+1$ innovation is introduced by the triangular part and another source of innovation is introduced at level $\nu$ from the sinusoidal term. The power of the test is studied for $L\geq \nu$. An example plot is given in Figure \ref{fig:lrt_innov}. 
We set the significance level of our test at $\alpha=0.05$, with $M = 1$ and $\lambda_{0} \in [1000,50000]$ as in the LRT for $J$-th level homogeneity. The empirical type 1 error and power plots from 10000 simulations are shown in Figure \ref{fig:lrt_innov} for $L=1$ (type 1 error in the absence of innovation) and $L=3,4$ and $5$ (power in the presence of innovation). We are interested in exploring the effects of the parameter $\lambda_0$ on the empirical type 1 error and power of the LRT for the absence of $L$-th level innovation. Again the empirical type 1 error lies close to the $5\%$ level as expected when the conditions of Theorem \ref{innov_chi2} are met. We also observe that the empirical power converges to 1 as the magnitude of the perturbation increases through the product $A\lambda_0$. Since the intensity model is still proportional to $\lambda_0$, this is also justified from Theorem \ref{innov_chi2} as the equivalent behavior is expected when $\lambda_0$ is fixed and $M$ increases towards infinity. Furthermore, it is noticeable that for a fixed $\lambda_0$, the power decreases as we increase $L$. This can be explained because increasing $L$ displaces the mass of the null distribution $\chi^2_{2^{L}}$ further to the right, making it harder for the test to distinguish between the null hypothesis and the true state of nature. %A conclusion of this study is that this statistical innovation detector based on Haar wavelets has a guaranteed rate of false positives but will be less efficient to detect the same innovation when it is located at finer scales.

%%%%%%%%%%%%%%%%%%%%%%%%%%%%%%%%%%%%%%%%%%%%%%%%%%%%%%%%%%%%%%%%%%%%%%%%%
%%%%%%%%%%%%%%%%%%%%%%%%%%%%%%%%%%%%%%%%%%%%%%%%%%%%%%%%%%%%%%%%%%%%%%%%%
%%%%%%%%%%%%%%%%%%%%%%%%%%%%%%%%%%%%%%%%%%%%%%%%%%%%%%%%%%%%%%%%%%%%%%%%%

\section{Statistical Thresholding}
\label{sec:thresh}
\setcounter{figure}{0}  
As stated in Section \ref{sec:wavest}, we can define a non-linear wavelet estimator of the intensity of a point process when a thresholding strategy is applied on the coefficient estimates. We initially define a general formulation for thresholding strategies in intensity estimation that we can adapt to different examples. To define a thresholding strategy, we need to choose a wavelet family for the estimation of the corresponding coefficients and a threshold operator that will be applied on the data. We consider  a collection of compactly supported mother wavelets $\left\{\psi_{L,k}, k\in \mathcal{K}_L \right\}$, where $\mathcal{K}_L$ is the ordered finite subset of $\mathbb{Z}$ containing the translation indexes that are used as a basis for $\widetilde{W_L}$, and further denote $K_L = |\mathcal{K}_L|$. For instance $\mathcal{K}_L=\left\{ 0,...,2^L-1 \right\}$ under the Haar basis if the intensity has support $[0,1)$ or $[0,T)$ with rescaled wavelets. Let $\{N_m,m=1,...,M\}$, $M\geq 1$, be a collection of independent realizations of the same point process $N$, we define $\bhB^L = (b_{m,i})\in \mathbb{R}^{M\times K_L}$, where $b_{m,i} \equiv \widehat{\beta}^{(m)}_{L,k_i}$ is the estimator of the true wavelet coefficient $\beta_{L,k_i}$ obtained from $N_m$.
\begin{figure}[t]
\centering
\[ \bhB^L =
\begin{pmatrix}
    \widehat{\beta}^{(1)}_{L,1}       & \widehat{\beta}^{(1)}_{L,2}  	& \widehat{\beta}^{(1)}_{L,3} 	& \widehat{\beta}^{(1)}_{L,4}\\
    \vdots 								    & \vdots   								& \vdots 								& \vdots \\
    \widehat{\beta}^{(m)}_{L,1}      &  \widehat{\beta}^{(m)}_{L,2}  	& \widehat{\beta}^{(m)}_{L,3}	& \widehat{\beta}^{(m)}_{L,4}\\
    \vdots 								    & \vdots	 								& \vdots 								& \vdots \\
    \widehat{\beta}^{(M)}_{L,1}      & \widehat{\beta}^{(M)}_{L,2} 	& \widehat{\beta}^{(M)}_{L,3}	& \widehat{\beta}^{(M)}_{L,4}
\end{pmatrix}
\Longrightarrow \bhT^L = \cT(\bhB^L) = 
\begin{pmatrix}
    \widehat{\beta}^{(1)}_{L,1}       & 0  										& \widehat{\beta}^{(1)}_{L,3} 	& 0\\
    \vdots 								    & \vdots   								& \vdots 								& \vdots \\
    \widehat{\beta}^{(m)}_{L,1}      & 0  										& \widehat{\beta}^{(m)}_{L,3}	& 0\\
    \vdots 								    & \vdots	 								& \vdots 								& \vdots \\
    \widehat{\beta}^{(M)}_{L,1}      & 0 										& \widehat{\beta}^{(M)}_{L,3}	& 0
\end{pmatrix}
\]
\caption{	\label{fig:example} Example output of a thresholding operator with $\mathcal{K}_L= \left\{1,2,3,4\right\}$.}
\end{figure}

We represent a thresholding operator $\cT: \mathbb{R}^{M\times K_L} \rightarrow \mathbb{R}^{M\times K_L}$ with $\bhT^L=\cT(\bhB^L)$ being the output where  each column of $\bhT^L$ is the corresponding column of $\bhB^L$ if a thresholding criterion $C$ is met, or a column of zeros if $C$ is not met (see illustration in Figure \ref{fig:example}). If the $i$-th column of $\bhB^L$ meets the criterion $C$ and is therefore kept by the operator $\cT$, then the estimator of $\beta_{L,k_i}$ used in the final reconstruction of $\lambda$ will be the sample mean $\frac{1}{M} \sum\limits_{m = 1}^M \widehat{\beta}^{(m)}_{L,k_i} $. A thresholding operator is applied between coarse and fine limits $j_0$ and $J$, respectively, resulting in a filtering of the information contained in the detail spaces $W_j, j_0 \leq j \leq J$. The effect of different choices for $j_0$ and $J$ is explored in Appendices \ref{sec:paramj0} and \ref{sec:paramJ}. Defining the $\mathbb{R}^{K_L}$ vector $\Psi_L(t) = (\psi_{L,k_1}(t),...,\psi_{L,k_{K_L}}(t))^T$, where $ k_1$ and $k_{K_L}$ are respectively the first and last elements of the index set $\mathcal{K}_L$, and $\bone_M = (1,...,1)^T$ the vector of ones of length $M$, the non-linear estimator can be formulated as
\begin{equation}
\label{rate_tresh}
\widehat{\lambda}^{J}_{\cT}(t) = \frac{1}{M}\sum\limits_{m = 1}^M\sum\limits_{k_i \in  \mathcal{K}_{j_0}}\widehat{\alpha}^{(m)}_{j_0,k_i}\phi_{j_0,k_i}(t)+
\frac{1}{M}\sum\limits_{L = j_0}^J\bone_M^T \bhT^L \Psi_L(t)\ .
\end{equation}

Similarly to the distinction made in \cite{Hardle1998} for density estimation, we define three procedures for thresholding. We are applying local thresholding if criterion $C$ considers each column of $\bhB^L$ separately, global thresholding if $C$ considers the entire matrix $\bhB^L$, and intermediate thresholding for other cases where $C$ considers subsets of columns. The criteria $C$ that we will propose here are based on variations of the previously defined $L$-th level innovation hypothesis test formulated in Section \ref{sec:lrtinnov}, and in doing so we assume that the conditions of Theorem \ref{innov_chi2} are always met for all $j_0 \leq L \leq J$. Our thresholding strategies hence take the form of multiple hypothesis testing procedures. It is consequently crucial to consider efficient ways of handling multiple hypothesis tests as ignoring this specificity could lead to a high number of truly zero coefficients to be kept in the reconstruction of $\lambda$.

When $M=1$, a common setting, $\bhB^L$ and $\bhT^L$ become row vectors with the $i$-th element of $\bhT^L$ being $\widehat{\beta}_{L,k_i}(1-\mathbbm{1}_{[-\delta_{k_i},\delta_{k_i}]}(\widehat{\beta}_{L,k_i}))$, where $\delta_{k_i}\geq 0$, $i=1,...,K_L$, are threshold levels that need to be chosen. De Miranda and Morettin \citeyearpar{DeMiranda2011} propose $\delta_{k_i} = \omega\sqrt{\var(\widehat{\beta}_{L,k_i})}$, with $\omega$ typically equal to 3. This requires a crude estimator of the variance of the coefficient estimators. The authors notice an equivalence between this method and using $\widehat{\beta}_{L,k_i}$ as a test statistic for the null hypothesis $\beta_{L,k_i} = 0$. This employs Chebyshev's inequality and works on the assumption that $\widehat{\beta}_{L,k_i}$ is approximately Gaussian. This parallel is interesting enough for us to use this thresholding operator as a comparison point in our simulations.

\subsection{Local thresholding with False Discovery Rate control}
Under this thresholding procedure we apply a hypothesis test to each coefficient with the null hypothesis being that this coefficient is zero. In the case of Haar wavelets, the LRT for $L$-th level innovation defined in Section \ref{sec:lrtinnov} can be reduced to the case of a single coefficient without any change to its asymptotic properties. 

Using a local thresholding operator with Haar wavelets requires a total of $Q = 2^{J+1}-2^{j0}$ hypothesis tests for coarse and fine resolution scales $j_0$ and $J$, respectively. For this thresholding scheme, the criterion $C$ considers individually the p-value of each test. A naive criterion $C$ is that the coefficient is kept if the p-value for the corresponding test is lower than some fixed significance level $\alpha$. However, in this case too few coefficients might be thresholded. The other approach that we explore here follows the statistical thresholding method of \cite{abramovich1995thresholding} which is based on the False Discovery Rate (FDR) defined in \cite{Benjamini1995}. Of the $Q$ hypotheses being tested, we say that $Q_0$ are true null hypotheses and the total number of rejected hypotheses is $R$, of which $F$ are falsely rejected. Note that $Q_0$ and $F$ are unknown quantities. The FDR is the expectation of the ratio $F/R$, and is the quantity we look to control. Since the FDR approach to multiple testing produced lower mean squared errors compared to the universal hard threshold for certain types of signals in \cite{abramovich1995thresholding}, it seems natural to carry it over to the Poisson intensity estimation model. This method positions itself between the naive approach where the error is only controlled at the very local level (coefficient-wise) and more constrained approaches like Bonferonni's correction where the error is instead simultaneously controlled among all tests (the family-wise error rate), with the latter being prone to power loss.

This procedure assumes independence of at least the $Q_0$ test statistics associated with the true null hypotheses. Under that setting the FDR is controlled by $\alpha$, a global significance level. Since our Poisson intensity estimation model introduces dependence (between scales) among the test statistics, \cite{Benjamini2001} demonstrate that a conservative modification of $\alpha$ to $\alpha_Q = \alpha/(\sum\limits_{i =1}^Q \frac{1}{i})$ allows us to extend the FDR control method for any joint distribution of the test statistics. The FDR is then bounded by $(Q_0/Q)\alpha$ which is lower than $\alpha$.  Now the thresholding procedure is as follows:
\begin{enumerate}
\item Determine the p-values $p_{L,k}$ of the LRT for each null hypothesis $H_0^{L,k}$:``$\beta_{L,k} = 0$'', for all $j_0 \leq L \leq J$ and $k\in\mathcal{K}_L$ and sort them by increasing value to obtain the ordered indexed set $\mathcal{P} = \left\{ p_1, \dots p_Q \right\}$, where $Q$ is the total number of tests considered in the thresholding range. Note that $Q$ does not depend on $M$.
\item For a given significance level $\alpha$, find the largest index $i$ that satisfies $p_i \leq (i/Q)\alpha_Q$ where $\alpha_Q = \alpha/(\sum\limits_{i =1}^Q \frac{1}{i})$. 
\item Criterion $C$ states that the coefficients corresponding to the p-values smaller than or equal to $p_i$ are kept.
\end{enumerate}
 
\subsection{Global thresholding with Holm-Bonferroni correction}
The global thresholding strategy is based on the exact $L$-th level innovation test defined in Section \ref{sec:lrtinnov}. In this circumstance we test each level $j$,  $j_0 \leq j \leq J$ with a single test. The total number of tests is now $Q = J-j_0+1$, significantly decreasing computational time when compared to the local thresholding method. Again, several approaches can be considered to control the multiplicity of errors arising from combining the results of multiple tests. One thing to notice is that swapping multiple univariate tests for a single multivariate test at each level $L$ is already a way to address multiple hypothesis testing in this context. This choice reflects an emphasis on the detection of any significant information inside the detail space $\widetilde{W_L}$ regardless of its temporal location. This makes the thresholding easier to control statistically but may lead to an unnecessary number of coefficients kept in the end. Now since the number of tests here is linear with the maximum resolution $J$ and thus limited in practice, the Holm-Bonferonni method, which is a uniformly more powerful method than Bonferonni correction, can be reasonably considered. Another interest here is that Holm-Bonferroni correction does not require independence of the test statistics. Now the procedure to determine the criterion $C$ is the following:
\begin{enumerate}
\item Determine the p-value of the LRT for each null hypothesis $H_0^L: $ `` there is no $L$-th level innovation'', $j_0\leq L \leq J$, and sort them by increasing value to obtain the ordered indexed set $\mathcal{P} = \left\{ p_1, \dots p_Q \right\}$, where $Q$ is the total number of tests considered in the thresholding range. Again $Q$ does not depend on $M$.
\item From a given significance level $\alpha$, find the minimal index $i$ that satisfies $p_i > \frac{\alpha}{Q+1-i}$. Note this index $i_m$.
\item Reject the null hypotheses with p-values indexed from $1$ to $i_m-1$.
\item Criterion $C$ states that if the test at level $L$ is rejected then $\bhT^L = 0$, otherwise  $\bhT^L = \bhB^L$.
\end{enumerate}
Using Holm-Bonferroni's correction, the familywise error rate of this global thresholding strategy, which is the probability or having at least one type 1 error for an individual test, is always less or equal to the given significance level $\alpha$.
\subsection{Intermediate thresholding based on recursive tests}

The intermediate thresholding strategy uses the recursive testing approach proposed in \cite{Ogden1996}. This method falls into the intermediate category since the number of coefficients tested together to determine Criterion $C$ varies between $1$ and $K_L = |\mathcal{K}_L|$ for each resolution level $L$. The procedure is the same at each level $j_0 \leq L \leq J$, and is as follows:
\begin{enumerate}
\item Test the null hypothesis $H_0^{L}: $``$\beta_{L,k} = 0$ for all $k \in \mathcal{K}_L$'' using the LRT at significance level $\alpha$.
\item If the test is rejected, find the index $i$ for which the sample mean $1/M \sum_m \widehat{\beta}^{(m)}_{L,i} $ has the largest absolute value. Remove the $i$-th component in the null hypothesis $H_0^L$ to form a new null hypothesis $H_0^{L,-i}$.
\item Repeat steps 1 and 2 until the null is not rejected. Criterion $C$ retains all the coefficients that have been removed from the original null hypothesis.
\end{enumerate}

\subsection{Simulation study}
\label{sec:sims}
This study aims to compare the accuracy of different thresholding strategies by applying them on three Poisson process models on $[0,1]$ with intensities that exhibit different behaviors and regularities. The chosen measure of accuracy is the root mean integrated squared error (RMISE) which we estimate with 
$$
\widehat{RMISE} = \frac{1}{n}\sum_{i=1}^n\left(\frac{1}{m}\sum_{j=1}^m\left(\widehat{\lambda}^J_i(t_j)-\lambda(t_j)\right)^2\right)^{1/2}.
$$
In these studies, we use $n=10000$ repeat simulations and $t_j = (j-1)/m$ where $m=1000$.
 The first two intensity models are based on the ``\textit{Blocks}" and ``\textit{Bumps}" test functions from \cite{Donoho1994}. The third function is a modification to that defined in Section \ref{sec:sim}.  We will refer to this model as ``\textit{TriangleSine}" and it has expression
$$
 f_{\textrm{tsine}}(t) = \lambda_0\left(\frac{2-\xi}{2} - s(t)(i(t)\bmod 2)\xi\right)  + s(t)a\left(t-\frac{i(t)}{2^{V+1}}T\right) + A\lambda_0\sin\left(\frac{2^{L+1}\pi}{T}t+\frac{1}{T}\right),$$
with $s(t) =  1 - 2(i(t)\bmod 2)$, $a = (2^{V+1}\xi\lambda_0)/T$,
and $i(t) \in \left\{0, \dots, 2^{V+1}-1\right\}$ is the index of the interval $[\frac{i(t)}{2^{V+1}}T,\frac{i(t)+1}{2^{V+1}}T]$ in which $t$ belongs. 

We set $T=1$ and rescale these functions so that their integral on $[0,1]$ are equal. Further, since the ``\textit{Blocks}" function can take negative values, we apply an upwards shift such that it is positive. The resulting intensities are 
\begin{equation*}
\resizebox{\textwidth}{!}{$
\lambda_{\textrm{blocks}}(t) = 1.75A_0 + 0.25A_0 \frac{f_{\textrm{blocks}}(t)}{\int\limits_0^1  f_{\textrm{blocks}}}\qquad
\lambda_{\textrm{bumps}}(t) = 1.75A_0 + 0.25A_0 \frac{f_{\textrm{bumps}}(t)}{\int\limits_0^1  f_{\textrm{bumps}}}\qquad
\lambda_{\textrm{tsine}}(t) = A_0 + A_0 \frac{f_{\textrm{tsine}}(t)}{\int\limits_0^1  f_{\textrm{tsine}}}. $
}\end{equation*}
\begin{figure}[t]
\centering
\includegraphics[width=\textwidth]{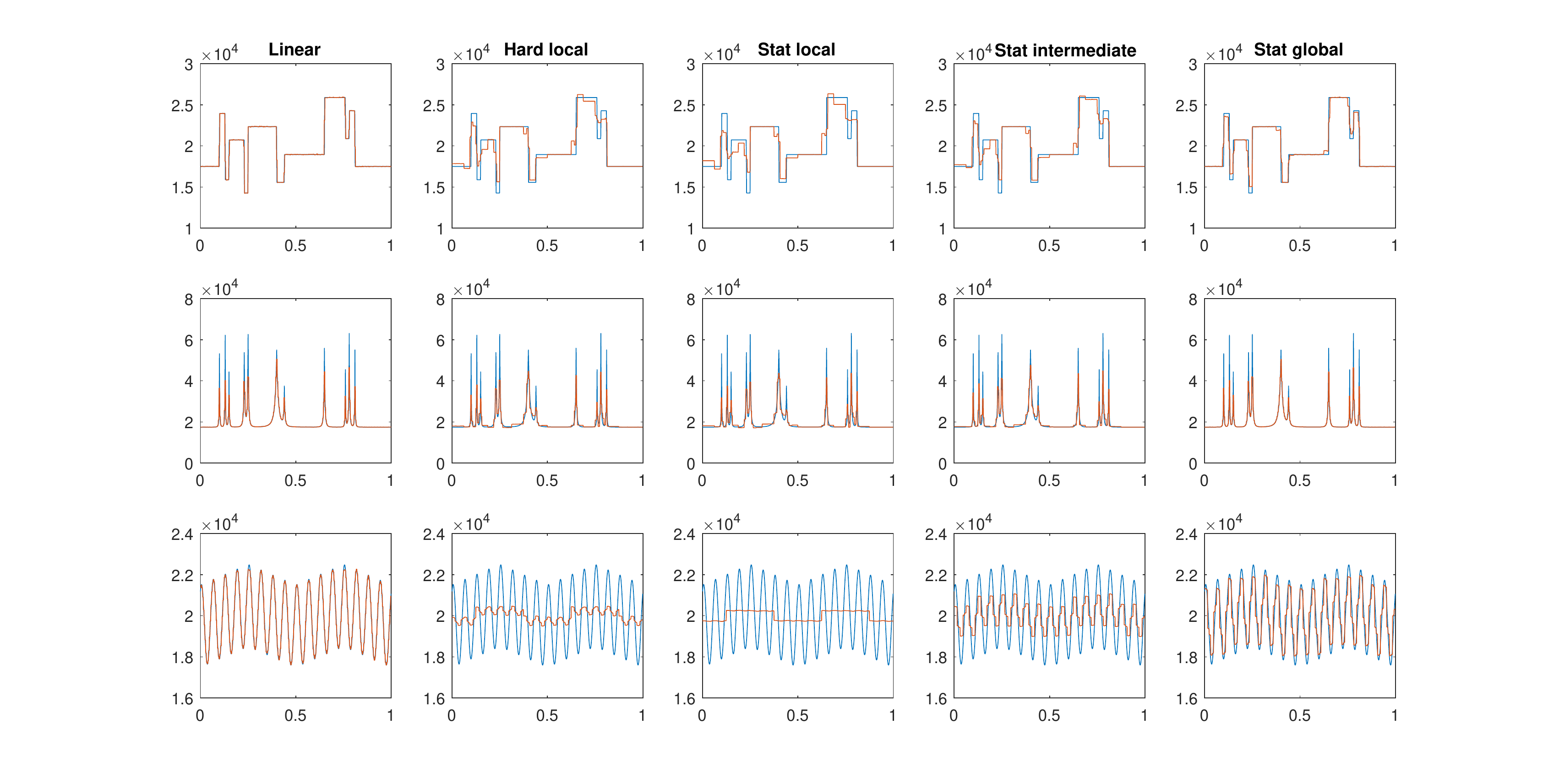}
\caption{Averaged reconstruction of the three intensity models ``\textit{Blocks}", ``\textit{Bumps}" and ``\textit{TriangleSine}", with $j_0 = 3, J = 7, M = 1$ and significance level $\alpha = 0.05$. The true intensity is in blue and the reconstruction is in red.}
\label{fig:thresh_sim}
\end{figure}
We are therefore ensuring that $E\{N(1)\} $ is always equal to $2A_0$ for the three Poisson process models. The value of $A_0$ determines the highest resolution at which we can threshold the Haar wavelet coefficients.From the conditions of Theorem \ref{innov_chi2}, we impose that the minimum value of the set $\left\{M\mu_{i} = M\int_{s^{J+1}_{i}}\lambda(t)\d t,\quad i=0,...,2^{J+1}-1 \right\}$ must be greater than or equal to 100 for reliable likelihood ratio tests for $L$-th level innovation up to level $J$ (and for smaller groups of wavelet coefficients in local and intermediate thresholding). Since we are demonstrating the presented methods for the $M=1$ case this imposes that the minimum value of $\{\mu_i, i=0,...,2^{J+1}-1\}$ is greater than or equal to 100. 

We now compare the RMISE on these three intensity models for five thresholding strategies: statistical local, intermediate and global thresholding, as well as no thresholding (linear estimation) and the hard local thresholding of \cite{DeMiranda2011}. We included the linear estimation as it serves as a reference point and is also the $M = 1$ case for the methods presented in \cite{Reynaud-Bouret2010} and \cite{Bigot2013}. We aim to study the influence of four parameters on this accuracy ranking: the starting resolution level $j_0$, the maximum resolution level  $J$, the significance level $\alpha$ and the value of $A_0$. In Table \ref{res-thresh5} we provide the relative RMISE (R-RMISE) values for one scenario where the estimated RMISE for each thresholding strategy is divided by the value under absence of thresholding, which serves as a reference point. We refer to the method of \cite{DeMiranda2011} as ``DM-L" and our three statistical thresholding strategies as ``LRT-L", ``LRT-I" and ``LRT-G" for the local, intermediate and global thresholding methods respectively. Intensity reconstructions averaged over 10000 simulations are shown in Figure \ref{fig:thresh_sim} under the same setting and for all thresholding procedures as well. Bootstrapped 95\% confidence intervals for the RMISE, plus further simulation studies can be found in Appendix \ref{apsec:sims}.
\begin{table}[t]
\centering
\begin{tabular}{l|l|l|l|l|l|}
\cline{2-6}
& \textbf{Linear}        & \textbf{DM-L}      & \textbf{LRT-L}     & \textbf{LRT-I}       & \textbf{LRT-G}             \\ \hline
\multicolumn{1}{|l|}{\textbf{Blocks}}       & 1               & 0.6455              & 0.6937              & \textbf{0.6402}            & 0.7701               \\ \hline
\multicolumn{1}{|l|}{\textbf{Bumps}}        & 1               & 1.0099              & 1.0538              & \textbf{0.9659}            & 0.9996               \\ \hline
\multicolumn{1}{|l|}{\textbf{TriangleSine}} & 1               & 0.6887              & 0.6544              & 0.6747                     & \textbf{0.6000}      \\ \hline
\end{tabular}
\caption{R-RMISE values with $j_0 = 3, J = 7, M = 1$ and significance level $\alpha = 0.05$. The number in bold indicates the best performing method.}
\label{res-thresh5}
\end{table}
The first conclusion in the setting of Table \ref{res-thresh5} is that we have statistical evidence that for all three intensity models at least one of LRT-I or LRT-G performs better than the linear and DM-L strategies. The statistical validity of this ranking relies on the absence of overlap between the 95\% confidence intervals for the RMISE of each method, as shown in Appendix \ref{apsec:sims} Table \ref{ci_thresh_j03}. LRT-G performs better when innovations are well spread across time, whereas LRT-I leads in the case of abrupt changes. This was expected from the design of each strategy. For instance, the ``\textit{Blocks}'' intensity has a sparse Haar wavelet decomposition with non-zero mother wavelets coefficients at high resolutions localized at the jumps. Therefore, this model favors LRT-L and LRT-I. Figure \ref{fig:thresh_sim} shows the mean intensity estimate against the true intensity and therefore illustrates bias. We note as expected that the linear estimator is unbiased, although it has high variance which is accounted for in the RMISE.

%%%%%%%%%%%%%%%%%%%%%%%%%%%%%%%%%%%%%%%%%%%%%%%%%%%%%%%%%%%%%%%%%%%%%%%%%
%%%%%%%%%%%%%%%%%%%%%%%%%%%%%%%%%%%%%%%%%%%%%%%%%%%%%%%%%%%%%%%%%%%%%%%%%
%%%%%%%%%%%%%%%%%%%%%%%%%%%%%%%%%%%%%%%%%%%%%%%%%%%%%%%%%%%%%%%%%%%%%%%%%

\section{Conclusion}
\label{sec:conc}
The wavelet analysis of point processes in continuous time has been addressed through wavelet expansions of the first-order intensity. By defining a multiresolution analysis on the point process, new multiscale properties, namely $J$-th level homogeneity and $L$-th level innovation, were introduced and tests for them formulated. Importantly, these tests can be applied when only a single realization of the process is observed. Tests for $L$-th level innovation formed the framework with which to perform thresholding of wavelet coefficients for intensity estimation. 

The root mean integrated squared error of these methods were compared on simulated data for three different intensity models, revealing different accuracy rankings depending on the model. An important point here is that no thresholding method uniformly outperforms all others - although at least one of the statistical thresholding (LRT) methods outperforms the existing local hard thresholding method (DM-L) in all but one of the scenarios studied (see Appendix \ref{apsec:sims}). This seems reasonable and is consistent with the study of \cite{Antoniadis2001} for wavelet regression and \cite{Besbeas2004} for discrete time Poisson intensity estimation. The rule of thumb we offer is that LRT-G outperforms the other methods for intensity functions that exhibit smooth, large-scale changes in time. For intensity functions that exhibit abrupt, localized changes (i.e. possess a sparse wavelet representation), LRT-L and LRT-I strategies are to be preferred. 

How to go about choosing the free-parameters $\alpha, j_0$ and $J$ in a data-driven way still needs to be addressed. The development of cross validation schemes in the point process setting would make an interesting extension but falls outside the scope of this paper. Extensions of the presented theory and methodology can now be considered for the second-order intensity and multidimensional point processes.

\appendix
\renewcommand\thefigure{\thesection.\arabic{figure}}
\renewcommand\thetable{\arabic{table}}
\section{Intensity Estimation with Daubechies D4 Wavelets}
\setcounter{figure}{0}  
\label{D2Q}
\subsection{Linear estimator}
\label{D2Qest}
The Daubechies D2Q wavelets \citep{Hardle1998} have $supp\ \phi \subseteq [0,2Q-1]$ and $supp\ \psi \subseteq [-Q+1,Q]$. When considering Daubechies D2Q wavelets with $Q > 1$, a closed form time domain approximation is needed as there does not exist an exact one. From a set of values obtained with the cascade algorithm \citep{Mallat1989}, we use a linear interpolation to approximate the mother and father wavelets.  As $supp\ \phi \subseteq [0,2Q-1]$, Daubechies D2Q wavelets do not have disjoint supports across all unit translations for a fixed scale. However $supp\ \phi$ is finite so we do have a finite number of coefficients that we estimate at each scale. For consistency between the different estimation methods, we desire that the interval $[0,T]$ coincide with the support of the Daubechies D2Q father wavelet at resolution 0. Taking $Q = 2$, this means rescaling the process $N$ to $[0,3]$, performing the estimation of its intensity, and rescaling this reconstruction back to $[0,T]$. We have the following linear estimator for the projection of the rescaled intensity onto $V_{J}$:
$$\widehat{\lambda}^{J}(t) = \sum\limits_{k = -2}^{(3 \times 2^{J})-1}\widehat{\alpha}_{J,k}\phi_{J,k}(t) .$$

\subsection{Coefficient-wise hypothesis test for local thresholding}
In order to define thresholding strategies we need to derive the distribution of the mother wavelet coefficients. Consider the collection of mother Daubechies D2Q wavelets $\left\{\psi_{L,k}, k \in \mathcal{K_L} \right\}$ that describes $\widetilde{W_{L}}$ at each scale $L$, where $\mathcal{K_L}$ is a finite set of indexes. Under the Daubechies D4 wavelet wavelet, we have $\mathcal{K_L} = \left\{ 1,...,(3 \times 2^L)-2 \right\} $ and hence $\widetilde{W_{L}}= {\rm span}\{\psi_{L,k};k=1,...,(3\times 2^L)-2\}$. As defined in Section \ref{sec:thresh}, let $\bhB^L = (b_{m,k})\in \mathbb{R}^{M\times K_L}$ where $M$ is the number of independent realizations of the point process $N$, $K_L = |\mathcal{K}_L|  = (3\times 2^L)-2$ and $b_{m,k} \equiv \widehat{\beta}^{(m)}_{L,k}$ is the estimator of the true wavelet coefficient $\beta_{L,k}$ obtained from $N_m$.

\begin{figure}[t]
\centering
\includegraphics[height=0.2\textheight]{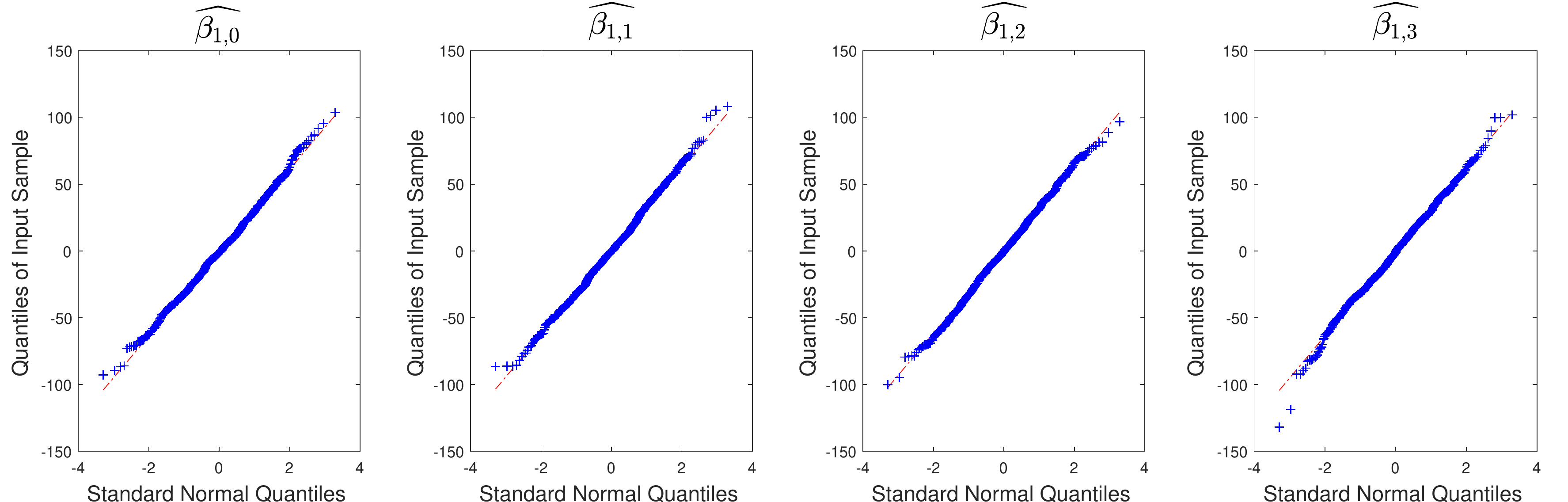}
\caption{QQ-plots for the coefficients estimates $\widehat{\beta}_{1,k}, k = 0,...,3$ with the Daubechies D4 wavelet and a homogeneous Poisson process with intensity $\lambda_0 = 1000$.}
\label{fig:qqplot}
\end{figure}

In order to extend the local thresholding scheme based on FDR control to Daubechies D4 wavelets, we need a hypothesis test for each single coefficient. The probability density function of the empirical coefficients for a compactly supported and continuous wavelet family is given in \cite{DeMiranda2008}. However if the wavelet is non tractable in time domain then so is its density. QQ-plots in Figure \ref{fig:qqplot} suggest that a Gaussian approximation is well suited when the coefficients are estimated using the stochastic integral $\widehat{\beta}_{j_0,k} = \int_{\mathbb{R}} \psi_{j,k}(t)\dif N(t) = \sum_{\tau_i\in\mathcal{E}}\psi_{j,k}(\tau_i)$ and $\psi$ is approximated as in \ref{D2Qest}. Also, a useful result from  \cite{DeMiranda2011} is $\widehat{\var}(\widehat{\beta}_{L,k}) = \int \psi_{L,k}^2(t)\dif N(t) = \sum_{\tau_i\in\mathcal{E}}\psi_{j,k}^2(\tau_i)$
is an unbiased estimator for the variance of coefficient $\widehat\beta_{L,k}$. With $M \geq 1$  independent realizations of the point process $N$, the estimator of $\beta_{L,k}$ used in the final reconstruction of $\lambda$ will be the sample mean $\frac{1}{M} \sum\limits_{m = 1}^M \widehat{\beta}^{(m)}_{L,k}$. Similarly, a variance estimator for  $\widehat\beta_{L,k}$ is  $\frac{1}{M} \sum\limits_{m = 1}^M \widehat{\var}(\widehat{\beta}^{(m)}_{L,k})$. Therefore, testing the hypothesis $H : \beta_{L,k} = 0$ against the alternative hypothesis $K :\beta_{L,k} \neq 0$ can be performed using  $\frac{1}{M} \sum\limits_{m = 1}^M \widehat{\beta}^{(m)}_{L,k}$ as a test statistic. Under the null hypothesis, we assume that $\widehat{\beta}^{(m)}_{L,k} \sim \mathcal{N}(0,\sigma_{L,k}^2)$. Since $\sigma_{L,k}^2$ is unknown we instead use $\frac{1}{M} \sum\limits_{m = 1}^M \widehat{\var}(\widehat{\beta}^{(m)}_{L,k})$ to estimate a confidence interval from a given significance level $\alpha$. All estimators are consistent so the approximate null distribution converges to the true null distribution as $M \rightarrow \infty$.
\subsection{$L$-th level innovation hypothesis test for global thresholding}
\begin{figure}[t]
\centering
\includegraphics[height=0.2\textheight]{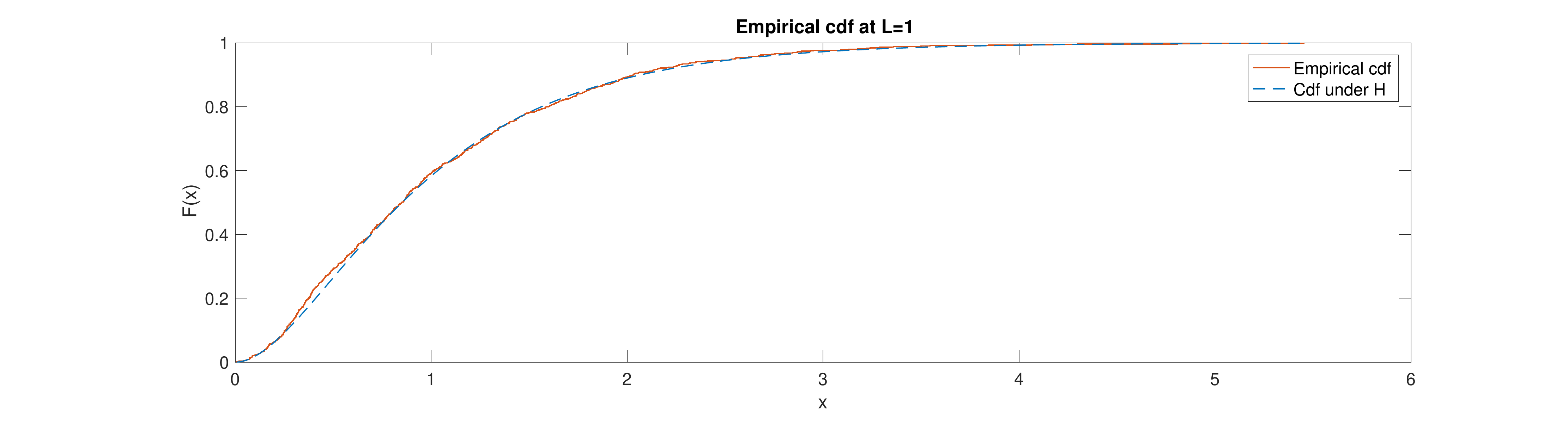}
\caption{Comparison between the empirical and null cumulative distribution functions of Hotelling's t-squared statistic with the Daubechies D4 wavelet and a homogeneous Poisson process at resolution $L = 1$ with $\lambda_0 = 1000$ and $M = 50$.}
\label{fig:ecdf_D4}
\end{figure}

We now want to design a multivariate test for the null  hypothesis $H : \mathit{\boldsymbol \mu_L = \mathbf{0}_L}$ where $\boldsymbol \mu_L$ is the mean vector of the coefficients $\widehat{\beta}_{L,k}, k \in \mathcal{K_L}$. Given the approximate normality of the coefficients estimates $\widehat{\beta}_{L,k}, k \in \mathcal{K_L}$ under the Daubechies D4 wavelet suggested in Figure \ref{fig:qqplot}, a possible choice of hypothesis test is the multivariate extension of the Student's t-test based on Hotelling's t-squared statistic. In our setting this statistic will be $t^{2} =(\bar{\boldsymbol \mu}_L)^T\widehat{\Sigma_L}\bar{\boldsymbol \mu}_L$ where $\bar{\boldsymbol \mu}_L$ is the sample mean of the empirical coefficients and $\widehat{\Sigma_L}$ their sample covariance. If the  estimators $\widehat{\beta}_{L,k}, k \in \mathcal{K_L}$ form a multivariate Gaussian vector, then under the null hypothesis $H$ this statistic is proportional to an F-distributed random variable with parameters $M$ and $K_L$. The empirical cumulative distribution function of $t^{2}$ shown in Figure \ref{fig:ecdf_D4} seems to follow closely the desired distribution under the null hypothesis. However, this particular hypothesis test requires that the sample size $M$ must always be greater than $K_L$, making it impossible to apply at higher resolutions for low values of $M$. We will therefore not develop this hypothesis test further in this work.

\section{Further Discussion on Likelihood Ratio Rests}
\setcounter{figure}{0}  
\label{LRTdiscussion}
\subsection{Null parameters on the boundary of the parameter space}
\label{ap:boundary}
In Theorems \ref{chi2} and \ref{innov_chi2}, we establish in particular the asymptotic distribution of the modified likelihood ratio statistic $R$ under $J$-th level homogeneity and $L$-th level innovation, respectively. These results require that the true but unknown values of the parameters $\lambda_c^J$ and $\mu_{k}^{pair}$ under the null hypothesis of each LRT are in the interior of the global parameter space. The boundaries of the parameter spaces when testing  $J$-th level homogeneity and $L$-th level innovation contain the parameter vectors satisfying $\lambda_k^J = 0$ and $\mu_{k} = 0$ for one or more dyadic translation indexes $k \in  0,...,2^J -1$ and $k \in 0,...,2^{L+1} -1$, respectively. If $\lambda_c^J = 0$ under $J$-th level homogeneity for any resolution level $J$, then we are in the trivial case where the intensity is the zero function on $[0,T)$. This would lead us to never observe any event almost surely, thus we can exclude the value $\lambda_c^J = 0$ from the null parameter space of our likelihood ratio model. From any other data, Theorem \ref{chi2} can be used provided its other conditions are met. 

However, point processes with non-zero intensities may not possess $L$-th level innovation for some level $L$ but still have parameters for that test on the boundary. For example, if $\lambda$ is non-zero and constant on $[0,T/2)$ and zero otherwise, then there is no level 1 innovation (in the Haar sense) but $\mu_1^{pair} = 0$. Therefore, we need to detail a further analysis in order to propose decision rules when $M$ is large and some MLEs of the parameters take value zero. This analysis will be done under the general setting of Section \ref{sec:lrt_pairs}. 

Under the null hypothesis of the model leading to Theorem \ref{innov_chi2}, we note $U$ the number of true parameters $\mu_{i}^{pair}$ equal to zero. By definition, we have $0 \leq U < P$, the case $U = P$ being excluded since under this condition no data points would be observed. From some data $\mathbb{X}$, we also note $\bar{U}$ the number of pairs of MLEs $(\bar{\mu}_{2i-1},\bar{\mu}_{2i})$ that are equal to $(0,0)$, which is equivalent to $\bar{\mu}_{i}^{pair} = 0$. If $\bar{U} = 0$, then $U=0$ since $U \leq \bar{U}$ and therefore the true parameter vector does not lie on the boundary of the parameter space, which allows us to apply Wilks' theorem provided its other conditions are met. Now consider the case where $\bar{U} > 0$ and thus $U > 0$ is a possibility. Let us place ourselves under the null hypothesis $H$, which we recall states $\mathit{\mu_{2i-1}=\mu_{2i} = \mu_i^{pair},\ 1 \leq i \leq P}$. From the proof in Appendix \ref{proof_Rstat_innov}, statistic $R$ would have the same value if the data $\mathbb{X}$ was instead obtained from a multivariate Poisson random variable of dimension $2P-2U$ where we exclude the $U$ pairs of components that have zero mean. Therefore, under the conditions of Theorem \ref{innov_chi2}, $R$ is asymptotically $\chi^{2}_{P-U}$ distributed. 

Since the value of $U$ is hidden and $U \leq \bar{U}$, we have $\bar{U}+1$ possible distributions for $R$ under the null and hence $\bar U+1$ possible critical values. Each critical value is noted $z_{u,\alpha}$ and is the upper $100(1-\alpha)\%$ point of the $\chi^{2}_{u}$ distribution. We propose three choices of critical values which yield different type 1 error bounds for the LRT. The asymptotic type 1 error of the LRT is noted $\epsilon_1$ and the cumulative distribution function of the chi-squared distribution with $d$ degrees of freedom is noted $\F_d$.  
\begin{enumerate}
\item The first choice is to use the critical value $z_{P,\alpha}$, which is equivalent to assuming $U=0$, i.e. all mean parameters $\mu_{i}^{pair}$ are non-zero. This places us in the most conservative setting since $z_{P,\alpha} = \max\left\{ z_{u,\alpha},    \ 0 \leq u \leq \bar{U} \right\}$. The type 1 error of the LRT in this case satisfies $ 1 - \F_{P-\bar{U}}(z_{P,\alpha}) \leq \epsilon_1 \leq \alpha$. This reduction in type 1 error is accompanied by a loss of power.
\item The second choice is to use the critical value $z_{P-\bar{U},\alpha}$, which is equivalent to assuming $U = \bar U$, i.e $\mu_{i}^{pair}=0$ if and only if $\bar \mu_i^{pair}=0$. Therefore, this is the maximum likelihood decision. It  leads to a gain of power when $U < \bar{U}$ since $z_{P-\bar{U},\alpha}=\min\left\{ z_{u,\alpha},    \ 0 \leq u \leq \bar{U} \right\}$. However, the type 1 error of the LRT in this case now satisfies $ \alpha  \leq \epsilon_1 \leq 1 - \F_{P}(z_{P-\bar{U},\alpha})$.
\item The third choice of critical value is motivated by an attempt to strike a balance between $z_{P,\alpha}$ and $z_{P-\bar{U},\alpha}$. We propose the intermediate value $z_{P-\lc \frac{\bar{U}}{2} \rc,\alpha}$. This provides a scheme for balancing the type 1 error/power trade-off. The type 1 error now satisfies  $  1 - \F_{P-\bar{U}}(z_{P-\lc \frac{\bar{U}}{2} \rc,\alpha})  \leq \epsilon_1 \leq 1 - \F_{P}(z_{P-\lc \frac{\bar{U}}{2} \rc,\alpha})$.
\end{enumerate}

\subsection{Maximizing the $J$-th level homogeneity test statistic}
\label{nopoints}
For a point process that is not level $J$ homogeneous, and for large $ M $, we can still encounter situations where one or several MLEs $\bar{\lambda}_k^J$ are equal to zero, whether the corresponding true parameters $\lambda_k^J$ are zero or not. Considering the general setting of the LRT in Section \ref{sec:lrt_scaled}, we derive the situation under which $R$, the test statistic defined in Proposition \ref{prop_Rstat}, is maximized.
\begin{proposition}
\label{prop_maxR}
Let $c > 0$, $P \geq 1$ and $f^P : (x_1,...,x_P) \mapsto \sum\limits_{i = 1}^{P}x_i\log(\frac{x_i}{c})$. Let $\Omega_{c,P}$ be the subset of $[0,Pc]^P$ defined as $\Omega_{c,P} = \left\{(x_1,...,x_P) \in [0,Pc]^P, \frac{1}{P}\sum\limits_{i=1}^{P}x_i = c \right\}$. Then the restriction of $f^P$ on $\Omega_{c,P}$ attains its maximum for any element in $\Omega_{c,P}$  of the form $(0,...,x_i = Pc,...,0), 1\leq i\leq P$.
\end{proposition}
See proof in Appendix \ref{proof_maxR}. In our setting, Proposition \ref{prop_maxR} has an interesting interpretation. If we impose that the MLE $\bar{\mu}_{c}$ takes some value $c > 0$, then statistic $R$ is maximized by the data $\mathbb{X}$ that produces one MLE $\bar{\mu}_{i}$ with value $Pc$ and all other MLEs with value zero. Proposition \ref{prop_maxR} illustrates a scenario of maximum inhomogeneity in the likelihood ratio sense, which is characterized by a maximum distance between the lowest and highest values among the MLEs $\bar{\mu}_{i}, 1 \leq i \leq P$. A similar result can be formulated for $L$-th level innovation.

\section{Proofs}
\setcounter{figure}{0}  
\label{proofs}
\subsection{Proof of Remark \ref{remark_haar}}
\label{proof_rem1}W.l.o.g. we prove this result with $T = 1$. Since $\alpha_{J,k} = \langle \lambda, \phi_{J,k} \rangle$, we have:
\begin{align*}
\alpha_{J,k} = \int\limits_0^1 \lambda(t)\phi_{J,k}(t)dt&= \int\limits_{k/2^J}^{(k+1)/2^J}2^{J/2} \lambda(t)\dif t \\
&= \frac{1}{\sqrt{2}}\left[\int\limits_{2k/2^{J+1}}^{(2k+1)/2^{J+1}} 2^{(J+1)/2}\lambda(t)\dif t + \int\limits_{(2k+1)/2^{J+1}}^{(2k+2)/2^{J+1}} 2^{(J+1)/2}\lambda(t)\dif t \right] \\
&= \frac{1}{\sqrt{2}}(\alpha_{J+1,2k} +\alpha_{J+1,2k+1} ).
\end{align*}

\subsection{Proof of Proposition \ref{prop}}
\label{proof_prop}
From the multiresolution setting defined in Section \ref{sec:dist_lambdaest}, we know that  $\widehat{\lambda}^{J}(t) = \widehat\lambda_k^J\cdot\mathbbm{1}_{s_k^J}(t)$ where $\widehat{\lambda}_{k}^{J}$ is the value of the $J$-th level wavelet reconstruction estimator on the subinterval $s^{J}_{k}\in S_J$. Using $\widehat{\alpha}_{J,k} =  \sum_{\tau_{i}}\phi_{J,k}(\tau_{i})$, for Haar wavelets we have $\widehat{\alpha}_{J,k} = \frac{2^{J/2}}{\sqrt{T}}x_k^J$, where $x_{k}^J$ is the event count in the corresponding subinterval $s^{J}_{k} \in S_J$. Therefore  $\widehat{\lambda}_{k}^{J} = \frac{2^{J}}{T}x_k^J$.  Since $N$ is a Poisson process and Haar wavelets have disjoint supports across all translations for a fixed scale $J$, we have:
\begin{enumerate}
\item each event count $x_{k}^J$  is Poisson distributed with mean $\mu_{k}^J = \int\limits_{s_{k}^J}\lambda(t)\d t$,
\item all event counts $x_{k}^J,\ 0 \leq k \leq 2^J-1,$ are independent.
\end{enumerate} 
Therefore the intensity estimators $\widehat\lambda_0^J,...,\widehat\lambda_{2^J -1}^J$ are independent random variables distributed as $$ \widehat{\lambda}^{J}_{k} \sim \frac{2^{J}}{T}\text{\rm Pois}(\mu^{J}_{k}).$$

\subsection{Proof of Proposition \ref{homo_eq}}
\label{proof_homo_eq}
\textit{Left to right:} This direction is proved using the definition of Haar wavelet coefficients. We know there exists $\lambda_0 \geq 0$ such that $\lambda(t) = \lambda_0$ a.e. (in the Lebesgue sense) on $[0,T)$. Let $J \geq 0$ and consider the subdivision $S_J= \left\lbrace s_{k}\right\rbrace_{k=0}^{2^{J}-1}$ defined in Section \ref{sec:dist_lambdaest}. For all $0 \leq k \leq 2^J -1$ we have:
$$
\alpha_{J,k} = \int\limits_{0}^T\lambda(t)\phi_{J,k}(t)\dif t 
= \int\limits_{s_k}\lambda(t)\phi_{J,k}(t)\dif t
=  \int\limits_{s_k}\lambda_0\phi_{J,k}(t)\dif t = \frac{2^{J/2}}{\sqrt{T}}\lambda_0.
$$
Hence for all $0 \leq k \leq 2^J -1$ and $0 \leq k' \leq 2^J -1$ we obtain $\alpha_{J,k} = \alpha_{J,k'}$. This is equivalent to $\lambda_k^J=\lambda_{k'}^J$ and $N \in H_J$.\\\\
\textit{Right to left:} To prove this direction we will consider the contrapositive. It is trivial that since $\lambda$ is piecewise continuous on  $[0,T)$, it is constant a.e. if and only if $\lambda(t) = \lambda_0  = \frac{1}{T}\int\limits_0^T \lambda(t)\dif t$ a.e on  $[0,T)$. Now, if $\lambda$ is not a function that is a.e. constant on $[0,T)$, then there exists two disjoint open intervals $I^{+}$ and $I^{-}$ in $[0,T)$ with non-zero length such that $\lambda$ is continuous on $I^{+}$ and $I^{-}$ as well as $\lambda_{|I^{+}} > \lambda_0$ and $\lambda_{|I^{-}} < \lambda_0$ where $\lambda_{|D}$ is the restriction of $\lambda$ on the interval $D$.
Let $d = \min(l(I^{+}),l(I^{-}))$, where $l(D)$ is the length of an interval $D$, and \[j_d = \argmin_j \frac{T}{2^j} < \frac{d}{2}. \]
Then there exists two indexes $k,k' \in \{0,...,2^{j_d} -1\}$, such that $[T\frac{k}{2^{j_d}},T\frac{k+1}{2^{j_d}}) \subset I^{+},\
[T\frac{k'}{2^{j_d}},T\frac{k'+1}{2^{j_d}}) \subset I^{-}$, and $\alpha_{j_d,k} > \alpha_{j_d,k'}$ since $\lambda$ is positive. This is equivalent to $\lambda_k^{j_d} > \lambda_{k'}^{j_d}$, which implies $N$ is not level $j_d$ homogeneous.

\subsection{Proof of Proposition \ref{prop_Rstat}}
\label{proof_Rstat}
Let $\mathbb{X} = \left\{\mathbf{X}_{m}\right\}_{m = 1}^{M}$ be a set of iid scaled Poisson random vectors, each with independent components of form $\mathbf{X}_{m} = \left(X_{m,i}\right)_{i = 1}^{P}$, $X_{m,i} \sim \delta \rm{Pois}(\mu_i)$. Therefore, for any non-negative integer $k_i$ we have $P(X_{m,i} = \delta k_{m,i} ) = \exp(-\mu_i)\frac{\mu_i^{k_{m,i} }}{k_{m,i} !}$. The likelihood functions of $\mathbb{X}$ under the null and alternative hypotheses $H$ and $K$ are
\begin{align*}
\L_H(\mathbb{X};\mu_{c},...,\mu_c) &=  \prod\limits_{m= 1}^{M}\prod\limits_{i = 1}^{P}\exp(-\mu_c)\frac{\mu_c^{k_{m,i}}}{k_{m,i}!} = \exp(-MP\mu_c)\prod\limits_{i = 1}^{P}\frac{\mu_c^{\sum\limits_{m= 1}^{M}k_{m,i} }}{\prod\limits_{m= 1}^{M} k_{m,i} !}, \\
\textrm{ and } \L_K(\mathbb{X};\mu_1,...,\mu_P) &=  \prod\limits_{m = 1}^{M}\prod\limits_{i = 1}^{P}\exp(-\mu_i)\frac{\mu_i^{k_{m,i} }}{k_{m,i} !} = \exp(-M\sum\limits_{i = 1}^{P}\mu_i)\prod\limits_{i = 1}^{P}\frac{\mu_i^{\sum\limits_{m= 1}^{M}k_{m,i} }}{\prod\limits_{m= 1}^{M}k_{m,i} !}.
\end{align*}
To locate their maxima we consider the log-likelihood functions
\begin{align*}
\log\L_H(\mathbb{X};\mu_{c},...,\mu_c) &=  -MP\mu_c + \sum\limits_{i = 1}^{P}\left[\log(\mu_c)\sum\limits_{m= 1}^{M}k_{m,i}  - \sum\limits_{m= 1}^{M} \log(k_{m,i} !) \right], \\
\textrm{ and } \log\L_K(\mathbb{X};\mu_1,...,\mu_P) &=   -M\sum\limits_{i = 1}^{P}\mu_i + \sum\limits_{i = 1}^{P}\left[\log(\mu_i)\sum\limits_{m= 1}^{M}k_{m,i}  - \sum\limits_{m= 1}^{M} \log(k_{m,i} !) \right].
\end{align*}
Differentiating each function with respect to its parameters gives:
\[
\begin{cases}
\dfrac{d\log\L_H}{d\mu_c} &= -MP + \frac{1}{\mu_c}\sum\limits_{i = 1}^{P}\sum\limits_{m= 1}^{M}k_{m,i}  \\
\dfrac{\partial\log\L_K}{\partial\mu_i} &= -M + \frac{1}{\mu_i}\sum\limits_{m= 1}^{M}k_{m,i},\ \forall \ 1\leq i \leq P.
\end{cases}
\]
Therefore, with $k_{m,i} = X_{m,i}/\delta$, the maximum values of $\L_H$ and $\L_K$ are respectively attained at
$\bar{\mu}_{c} = \frac{1}{\delta MP}\sum\limits_{i = 1}^{P}\sum\limits_{m= 1}^{M}X_{m,i}$ and  $\bar{\mu}_{i}= \frac{1}{\delta M}\sum\limits_{m= 1}^{M}X_{m,i}$ for all $1\leq i \leq P$. Statistic $\bar{\mu}_{c}$ is the MLE of $\mu_c$, the constant intensity under the null hypothesis $H$, and $\bar{\mu}_{i}$ is the MLE for $\mu_{i}$ ($i=1,...,P$) under the alternative hypothesis $K$. Since the likelihood ratio statistic $r$ is
\[
r =  \quad  \frac{\quad \mysup{} {\mu_{c} > 0}\ \L(\mathbb{X};\mu_{c},...,\mu_c) \quad }{\quad \mysup{}{\left\{\mu_{i}\right\}_{i = 1}^{P}, \sum \mu_i > 0}\ \L(\mathbb{X};\mu_1,...,\mu_P) \quad },
\]
applying the previous results yields
\[
r = \exp\left(-M\left(P\bar{\mu}_{c}-\sum\limits_{i = 1}^{P}\bar{\mu}_{i}\right)\right)\prod\limits_{i = 1}^{P}\left(\frac{\bar{\mu}_{c}}{\bar{\mu}_{i}}\right)^{M\bar{\mu}_{i}}
= \prod\limits_{i = 1}^{P}\left(\frac{\bar{\mu}_{c}} {\bar{\mu}_{i}}\right)^{M\bar{\mu}_{i}}.
\]
We can now derive the test statistic $R$:
$$R=-2\log\left(r\right)=2M\sum\limits_{i = 1}^{P}\bar{\mu}_{i}\log\left(\frac{\bar{\mu}_{i}}{\bar{\mu}_{c}}\right).$$

\subsection{Proof of Proposition \ref{prop_Rstat_innov}}
\label{proof_Rstat_innov}

Let $\mathbb{X} = \left\{\mathbf{X}_{m}\right\}_{m = 1}^{M}$ be a set of iid Poisson random vectors, each with independent components of form $\mathbf{X}_{m} = \left(X_{m,i}\right)_{i = 1}^{2P}$, $X_{m,i} \sim \rm{Pois}(\mu_i)$.  Therefore, for any non-negative integer $k_i$ we have $P(X_{m,i} =  k_{m,i} ) = \exp(-\mu_i)\frac{\mu_i^{k_{m,i} }}{k_{m,i} !}$. The likelihood functions of $\mathbb{X}$ under the null and alternative hypotheses $H$ and $K$ are
\begin{align*}
\L_H(\mathbb{X};\mu^{pair}_1,\mu^{pair}_1,...,\mu^{pair}_P,\mu^{pair}_P) &=  \prod\limits_{m= 1}^{M}\prod\limits_{i = 1}^{P}\exp(-2\mu^{pair}_i)\frac{(\mu^{pair}_i)^{k_{m,2i-1}+k_{m,2i}}}{k_{m,2i-1}!k_{m,2i}!} \\
&= \exp(-2M\sum\limits_{i = 1}^{P}\mu^{pair}_i)\prod\limits_{i = 1}^{P}\frac{(\mu^{pair}_i)^{\sum\limits_{m= 1}^{M}k_{m,2i-1}+k_{m,2i} }}{\prod\limits_{m= 1}^{M} k_{m,2i-1}!k_{m,2i}!}, \\
\textrm{ and } \L_K(\mathbb{X};\mu_1,...,\mu_{2P})&=  \prod\limits_{m = 1}^{M}\prod\limits_{i = 1}^{2P}\exp(-\mu_i)\frac{\mu_i^{k_{m,i} }}{k_{m,i} !} \\
&= \exp(-M\sum\limits_{i = 1}^{2P}\mu_i)\prod\limits_{i = 1}^{2P}\frac{\mu_i^{\sum\limits_{m= 1}^{M}k_{m,i} }}{\prod\limits_{m= 1}^{M}k_{m,i} !}.
\end{align*}
Then similarly as in \ref{proof_Rstat}, the likelihood function $\L_H$ is maximized when each parameter $\mu^{pair}_i$ is equal to $\bar{\mu}_{i}^{pair}= \frac{1}{2M}\sum\limits_{m= 1}^{M}k_{m,2i-1}+k_{m,2i} $, and the likelihood function $\L_K$ is maximized when each parameter $\mu_i$ is equal to $\bar{\mu}_{i}= \frac{1}{M}\sum\limits_{m= 1}^{M}k_{m,i}$. We also immediately have $\bar{\mu}_{i}^{pair} = \frac{1}{2}(\bar{\mu}_{2i-1}+\bar{\mu}_{2i})$. Since the likelihood ratio statistic $r$ is
\begin{align*}
&r =  \frac{\hspace{1.5cm} \mysup{} {\left\{\mu_{i}^{pair}\right\}_{i = 1}^{P},\ \sum \mu_i^{pair} > 0}\ \L(\mathbb{X};\mu^{pair}_{1},...,\mu^{pair}_{P}) \quad }{\hspace{1.5cm} \mysup{}{\left\{\mu_{i}\right\}_{i = 1}^{2P},\ \sum \mu_i > 0}\ \L(\mathbb{X};\mu_1,...,\mu_{2P}) \quad },
\end{align*}
applying the previous results yields
\begin{align*}
r &= \exp\left(-M\left(2\sum\limits_{i = 1}^{P}\bar{\mu}_{i}^{pair}-\sum\limits_{i = 1}^{2P}\bar{\mu}_{i}\right)\right)\prod\limits_{i = 1}^{P}\dfrac{\left(\bar{\mu}_{i}^{pair}\right)^{2M\bar{\mu}_{i}^{pair}}}{\left(\bar{\mu}_{2i-1}\right)^{M\bar{\mu}_{2i-1}}\left(\bar{\mu}_{2i}\right)^{M\bar{\mu}_{2i}}} \\
&= \prod\limits_{i = 1}^{P}\dfrac{\left(\bar{\mu}_{i}^{pair}\right)^{2M\bar{\mu}_{i}^{pair}}}{\left(\bar{\mu}_{2i-1}\right)^{M\bar{\mu}_{2i-1}}\left(\bar{\mu}_{2i}\right)^{M\bar{\mu}_{2i}}} \\
& =  \prod\limits_{i = 1}^{P}\dfrac{\left(\bar{\mu}_{i}^{pair}\right)^{M\bar{\mu}_{2i-1}}{\left(\bar{\mu}_{i}^{pair}\right)^{M\bar{\mu}_{2i}}}}{\left(\bar{\mu}_{2i-1}\right)^{M\bar{\mu}_{2i-1}}\left(\bar{\mu}_{2i}\right)^{M\bar{\mu}_{2i}}}.
\end{align*}
We can now derive the test statistic $R$:
$$R=-2\log\left(r\right)=2M\left[\sum\limits_{i = 1}^{P}\bar{\mu}_{2i-1}\log\left(\frac{\bar{\mu}_{2i-1}}{\bar{\mu}_{i}^{pair}}\right)+\sum\limits_{i = 1}^{P}\bar{\mu}_{2i}\log\left(\frac{\bar{\mu}_{2i}}{\bar{\mu}_{i}^{pair}}\right)\right].$$

\subsection{Proof of Theorem \ref{chi2}}
\label{proof_chi2}

The expression of $R$ given in Proposition \ref{prop_Rstat} can be rewritten as
$$R=2\sum\limits_{i = 1}^{P}\left[\sum_{m=1}^M \frac{X_{m,i}}{\delta}\log\left(\frac{P\sum_{m=1}^M X_{m,i}}{{\sum_{j = 1}^{P}\sum_{m=1}^M X_{m,j}}}\right)\right],$$
when replacing the MLEs by their actual value. With notation $Y_{i}^M=\sum_{m=1}^M X_{m,i}/\delta $, this becomes
 $$R=2\sum\limits_{i = 1}^{P}\left[Y_{i}^M\log\left(\frac{PY_{i}^M}{\sum_{j = 1}^{P}Y_{j}^M}\right)\right].$$
Given $Y_{i}^M$ is Poisson distributed with mean $\mu_{c}M$ under the null hypothesis $H$ ($\mu_{i}M$ under the alternative hypothesis $K$), the distribution of $R$ depends only on the product $\mu_{c}M$ (or $\mu_{i}M$). Therefore, the standard asymptotic results for $R$ hold as $\mu_c M\rightarrow \infty$. This limit can be achieved either through $M\rightarrow\infty$, $\mu_c\rightarrow\infty$, or both. The null distribution of $R$ is asymptotically $\chi^{2}$ with $P-1$ degrees of freedom for a large $\mu_c M$. We thus reject $H$ at significance level $\alpha$ if $R>c_\alpha$ where $c_{\alpha}$, the critical value, is the upper $100(1-\alpha)\%$ point of the $\chi^{2}_{P- 1}$ distribution. 

\subsection{Proof of Theorem \ref{innov_chi2}}
\label{proof_chi2_innov}

Similarly as in the proof for  Theorem \ref{chi2}, we go back to the expression of $R$ given in Proposition \ref{prop_Rstat_innov}:
$$R=2M\left[\sum\limits_{i = 1}^{P}\bar{\mu}_{2i-1}\log\left(\frac{\bar{\mu}_{2i-1}}{\bar{\mu}_{i}^{pair}}\right)+\sum\limits_{i = 1}^{P}\bar{\mu}_{2i}\log\left(\frac{\bar{\mu}_{2i}}{\bar{\mu}_{i}^{pair}}\right)\right].$$
This can also be written as
\begin{equation*}
\resizebox{\textwidth}{!}{$R=2\left[\sum\limits_{i = 1}^{P}\sum_{m=1}^M X_{m,2i-1}\log\left(\frac{2\sum_{m=1}^M X_{m,2i-1}}{\sum_{m=1}^M X_{m,2i-1} + X_{m,2i} }\right)+\sum\limits_{i = 1}^{P}\sum_{m=1}^M X_{m,2i}\log\left(\frac{2\sum_{m=1}^M X_{m,2i}}{\sum_{m=1}^M X_{m,2i-1} + X_{m,2i} }\right)\right]$
}\end{equation*}
when replacing the MLEs by their actual value. With the notation $Y_{i}^M=\sum_{m=1}^M X_{m,i}$, this becomes
$$R = 2\left[\sum\limits_{i = 1}^{P}Y_{2i-1}^M\log\left(\frac{2Y_{2i-1}^M}{Y_{2i-1}^M + Y_{2i}^M}\right)+\sum\limits_{i = 1}^{P}Y_{2i}^M\log\left(\frac{2Y_{2i}^M}{Y_{2i-1}^M + Y_{2i}^M }\right)\right].$$
Given $Y_{i}^M$ is Poisson distributed with mean $\mu_{i}^{pair}M$ under the null hypothesis $H$ ($\mu_{i}M$ under the alternative hypothesis $K$), the distribution of $R$ depends only on the product $\mu_{i}^{pair}M$ (or $\mu_{i}M$). Therefore, the standard asymptotic results for $R$ hold as $\mu_{i}^{pair} M\rightarrow \infty$, for all $1\leq i \leq P$. This limit can be achieved either through $M\rightarrow\infty$, $\mu_i^{pair}\rightarrow\infty$ for all $1\leq i \leq P$, or both. The null distribution of $R$ is asymptotically $\chi^{2}$ with $P$ degrees of freedom for all $\mu_i^{pair} M$ large. We thus reject $H$ at significance level $\alpha$ if $R>c_\alpha$ where $c_{\alpha}$, the critical value, is the upper $100(1-\alpha)\%$ point of the $\chi^{2}_{P}$ distribution. 

\subsection{Proof of Proposition \ref{prop_maxR}}
\label{proof_maxR}
We prove this by mathematical induction on the number of parameters $P$. The result is obvious at $P=1$ since $\Omega_{c,1}$ becomes the singleton $\{ c \}$. We will therefore detail the case $P=2$.

\subsubsection*{Base case $P=2$ :}
We have $\Omega_{c,2} = \left\{ (x_1,x_2) \in [0,2c]^2, x_1 + x_2 = 2c\right\}$. This lets us write  \[f^2(x_1,x_2) = x_1\log(\frac{x_1}{c}) + x_2\log(\frac{x_2}{c}) = x_1\log(\frac{x_1}{c}) + (2c-x_1)\log(\frac{2c-x_1}{c}).\]
Noting $g : x_1 \mapsto  x_1\log(\frac{x_1}{c}) + (2c-x_1)\log(\frac{2c-x_1}{c})$, then $g$ is differentiable with respect to $x_1$ on $(0,2c)$ and for all $x_1 \in (0,2c)$ we have: \[g'(x_1) = \log(\frac{x_1}{c}) + 1 - \log(\frac{2c-x_1}{c}) - 1 = \log(\frac{x_1}{2c - x_1}).\]
Immediately, $g'(x_1) = 0$ when $x_1 = c$, $g'(x_1) \leq 0$ when $x_1 \leq c$ and $g'(x_1) \geq 0$ when $x_1 \geq c$. Hence $g$ attains a local minimum at $x_1 = c$ and $\max\limits_{x_1 \in [0,2c]}g(x_1) = g(0) = g(2c) = 2c\log(2)$. Similarly, the restriction of $f^2$ on $\Omega_{c,2}$ is minimized at $(x_1,x_2) = (c,c)$ and maximized at $(2c,0)$ and $(0,2c)$.
\subsubsection*{Inductive step:}
Assume $P \geq 2$ and the restriction of $f^P$ on $\Omega_{c,P}$ is maximized at any vector $(x_1,...,x_P)$ of the form $(0,...,x_i = Pc,...,0), 1\leq i\leq P$. Let $x_{P+1} \in [0,(P+1)c]$ and $c_{x_{P+1}} = \frac{1}{P}((P+1)c-x_{P+1})$. For any $(x_1,...,x_P) \in \Omega_{c_{x_{P+1}},P}$, we have $\sum\limits_{i=1}^{P}x_i + x_{P+1} = Pc_{x_{P+1}} + x_{P+1} = (P+1)c$, hence $(x_1,...,x_P,x_{P+1}) \in \Omega_{c,P+1}$. Since the converse is also true we have $\Omega_{c,P+1} = \bigcup\limits_{x_{P+1} \in [0,(P+1)c]} \Omega_{c_{x_{P+1}},P} \times  \{x_{P+1}\}$, where $A \times B$ is the cartesian product of the sets $A$ and $B$. We know from the initial assumption that with a fixed value of $x_{P+1}$ the restriction of $h^{P}: (x_1,...,x_P) \mapsto f^{P+1}((x_1,...,x_P),x_{P+1})$ on $\Omega_{c_{x_{P+1}},P}$ is maximized when  $(x_1,...,x_P)$ is a vector belonging to the set $\left\{(0,...,x_i = Pc_{x_{P+1}},...,0), 1\leq i\leq P\right\}$. We now want to find the values $\tilde{x}_{P+1}$ that satisfy:
\[  \tilde{x}_{P+1} = \argmax\limits_{x_{P+1} \in [0,(P+1)c]} \max\limits_{\Omega_{c_{x_{P+1}},P}} h^{P}(x_1,...,x_P). \]
Noting $g_i : x_{P+1} \mapsto f^{P+1}(0,...,x_i = Pc_{x_{P+1}},...,0,x_{P+1})$, then $g_i$ is differentiable with respect to $x_{P+1}$ on the open interval $(0,(P+1)c)$ and for all $x_{P+1} \in (0,(P+1)c)$ we have: \[g'_i(x_{P+1}) = \log\left(\frac{x_{P+1}}{c}\right) + 1 - \log\left(\frac{(P+1)c-x_{P+1}}{c}\right) - 1 = \log\left(\frac{x_{P+1}}{(P+1)c - x_{P+1}}\right).\]
Similarly as in the base case, $g_i$ attains a local minimum in the open interval $(0,(P+1)c)$ when $x_{P+1} = Pc$. It also attains a maximum on $[0,(P+1)c]$ at $x_{P+1} = 0$, giving $c_{x_{P+1}} = \frac{P+1}{P}c$, and at $x_{P+1} = (P+1)c$, giving $c_{x_{P+1}} = 0$. Therefore the restriction of $f^{P+1}$ on $\Omega_{c,P+1}$ is maximized when $(x_1,...,x_P,x_{P+1}) \in \left\{(0,...,x_i = (P+1)c,...,0), 1\leq i\leq P+1\right\}$.

\section{Influence of Different Parameters on the RIMSE}
\label{apsec:sims}
\setcounter{table}{0}  
\subsection{Influence of $\mathbf{j_0}$}
\label{sec:paramj0}
\begin{table}[h!]
\centering
\resizebox{\textwidth}{!}{%
\begin{tabular}{l|l|l|l|l|l|}
\cline{2-6}
 & \textbf{Linear} & \textbf{DM-L} & \textbf{LRT-L} & \textbf{LRT-I} & \textbf{LRT-G} \\ \hline
\multicolumn{1}{|l|}{\textbf{Blocks}} & 2317 ([2315,2319]) & 1495 ([1493,1497]) & 1607 ([1604,1609]) & \textbf{1483 ([1481,1486])} & 1782 ([1776,1788]) \\ \hline
\multicolumn{1}{|l|}{\textbf{Bumps}} & 3061 ([3059,3063]) & 3091 ([3089,3094]) & 3226 ([3223,3229]) & \textbf{2957 ([2954,2959])} & 3060 ([3058,3061]) \\ \hline
\multicolumn{1}{|l|}{\textbf{TriangleSine}} & 2267 ([2265,2269]) & 1561 ([1560,1563]) & 1484 ([1483,1484]) & 1530 ([1528,1531]) & \textbf{1360 ([1355,1365])} \\ \hline
\end{tabular}%
}
\caption{Bootstrapped 95\% confidence intervals for the RIMSE with $A_0 = 10000, j_0 = 3, J = 7, M = 1$ and significance level $\alpha = 0.05$. The number in bold indicates the best performing method for each intensity model.}
\label{ci_thresh_j03}
\end{table}

\begin{table}[h!]
\centering
\resizebox{\textwidth}{!}{%
\begin{tabular}{c|c|c|c|c|c|}
\cline{2-6}
                                            & \textbf{Linear}    & \textbf{DM-L}      & \textbf{LRT-L}     & \textbf{LRT-I}              & \textbf{LRT-G}              \\ \hline
\multicolumn{1}{|l|}{\textbf{Blocks}}       & 2317 ([2315,2319]) & 1530 ([1528,1533]) & 1637 ([1635,1640]) & \textbf{1493 ([1490,1495])} & 1770 ([1765,1776])          \\ \hline
\multicolumn{1}{|l|}{\textbf{Bumps}}        & 3059 ([3057,3060]) & 3125 ([3122,3128]) & 3241 ([3238,3244]) & \textbf{2971 ([2968,2973])} & 3065 ([3063,3066])          \\ \hline
\multicolumn{1}{|l|}{\textbf{TriangleSine}} & 2266 ([2264,2268]) & 1541 ([1540,1542]) & 1458 ([1457,1458]) & 1514 ([1513,1515])          & \textbf{1303 ([1299,1308])} \\ \hline
\end{tabular}%
}
\caption{Bootstrapped 95\% confidence intervals for the RIMSE with $A_0 = 10000, j_0 = 0, J = 7, M = 1$ and significance level $\alpha = 0.05$. The number in bold indicates the best performing method for each intensity model.}
\label{ci_thresh_j00}
\end{table}
Decreasing $j_0$ from 3 to 0 is only slightly beneficial for the \textit{TriangleSine} model and increases the RIMSE in the two other intensity models. The amelioration observed for \textit{TriangleSine} could be explained by the absence of innovation at levels $0$ and $1$, and therefore the truly zero coefficients from these scales are less likely to be kept.

\subsection{Influence of $J$ and $\mathbf{A}_0$}
\label{sec:paramJ}
\begin{table}[h!]
\centering
\resizebox{\textwidth}{!}{%
\begin{tabular}{l|l|l|l|l|l|}
\cline{2-6}
                                            & \textbf{Linear}       & \textbf{DM-L}               & \textbf{LRT-L}        & \textbf{LRT-I}                 & \textbf{LRT-G}              \\ \hline
\multicolumn{1}{|l|}{\textbf{Blocks}}       & 8798 ([8792,8803])    & \textbf{6896 ([6891,6902])} & 7037 ([7032,7043])    & 7116 ([7110,7123])             & 8795 ([8789,8800])          \\ \hline
\multicolumn{1}{|l|}{\textbf{Bumps}}        & 21812 ([21809,21815]) & 21585 ([21581,21588])       & 21610 ([21607,21614]) & \textbf{21516 ([21513,21519])} & 21812 ([21809,21815])       \\ \hline
\multicolumn{1}{|l|}{\textbf{TriangleSine}} & 7324 ([7317,7330])    & 8297 ([8288,8306])          & 8697 ([8687,8708])    & 6873 ([6866,6879])             & \textbf{6860 ([6845,6875])} \\ \hline
\end{tabular}%
}
\caption{Bootstrapped 95\% confidence intervals for the RIMSE with $A_0 = 100000, j_0 = 3, J = 7, M = 1, $ and significance level $\alpha = 0.05$. The number in bold indicates the best performing method for each intensity model.}
\label{ci_thresh_l100000_5}
\end{table}

\begin{table}[h!]
\centering
\resizebox{\textwidth}{!}{%
\begin{tabular}{l|l|l|l|l|l|}
\cline{2-6}
                                            & \textbf{Linear}       & \textbf{DM-L}               & \textbf{LRT-L}        & \textbf{LRT-I}                 & \textbf{LRT-G}        \\ \hline
\multicolumn{1}{|l|}{\textbf{Blocks}}       & 14435 ([14429,14441]) & \textbf{6371 ([6362,6380])} & 6840 ([6831,6849])    & 6661 ([6651,6671])             & 11216 ([11181,11252]) \\ \hline
\multicolumn{1}{|l|}{\textbf{Bumps}}        & 16220 ([16212,16229]) & 14805 ([14792,14818])       & 15780 ([15767,15793]) & \textbf{13562 ([13547,13576])} & 16217 ([16208,16225]) \\ \hline
\multicolumn{1}{|l|}{\textbf{TriangleSine}} & 14314 ([14308,14320]) & 8568 ([8558,8577])          & 10045 ([10033,10057]) & \textbf{6916 ([6908,6923])}    & 6987 ([6960,7015])    \\ \hline
\end{tabular}
}
\caption{Bootstrapped 95\% confidence intervals for the RIMSE with $A_0 = 100000, j_0 = 3, J = 9, M = 1, $ and significance level $\alpha = 0.05$. The number in bold indicates the best performing method for each intensity model.}
\label{ci_thresh_l100000_5_J9}
\end{table}

Here we increase the value of $A_0$ from 10000 to 100000, the effect of which is to increase the power of each individual LRT involved in the statistical thresholding strategies. For the \textit{Blocks} model, we observe that DM-L is performing better than LRT-L, LRT-I and LRT-G. A study on the asymptotic evolution of the RIMSE values as $A_0 \rightarrow \infty$ could be done to verify this change of ranking. We also look at the effect of increasing $J$ from 7 to 9 while fixing $A_0 = 100000$. This leads to a significant decrease of the RIMSE for all thresholding strategies in the \textit{Bumps} model, as the peaks are located at very fine scales. As expected, it also increases the RIMSE for Linear and LRT-G under the \textit{Blocks} model as they keep a larger number of unnecessary coefficients, whereas the performance of DM-L, LRT-L and LRT-I is improved with this choice. However, a significant increase is observed for all thresholding strategies with the \textit{TriangleSine} intensity, which indicates that high resolutions terms penalize the RIMSE in this model.

\subsection{Influence of $\boldsymbol \alpha$}

\begin{table}[h!]
\centering
\resizebox{\textwidth}{!}{%
\begin{tabular}{l|l|l|l|l|l|l|}
\cline{2-7}
                                            & \textbf{LRT-L, $\alpha$ = 0.01} & \textbf{LRT-L, $\alpha$ = 0.05} & \textbf{LRT-I, $\alpha$ = 0.01} & \textbf{LRT-I, $\alpha$ = 0.05} & \textbf{LRT-G, $\alpha$ = 0.01} & \textbf{LRT-G, $\alpha$ = 0.05} \\ \hline
\multicolumn{1}{|l|}{\textbf{Blocks}}       & 1694 ([1692,1697])              & \textbf{1607 ([1604,1609])}     & 1530 ([1529,1532])              & \textbf{1483 ([1481,1485])}     & \textbf{1648 ([1644,1652])}     & 1782 ([1776,1788])              \\ \hline
\multicolumn{1}{|l|}{\textbf{Bumps}}        & 3475 ([3472,3479])              & \textbf{3226 ([3223,3229])}     & 3090 ([3088,3093])              & \textbf{2957 ([2954,2959])}     & \textbf{3060 ([3058,3061])}     & 3060 ([3058,3061])              \\ \hline
\multicolumn{1}{|l|}{\textbf{TriangleSine}} & \textbf{1480 ([1479,1480])}     & 1484 ([1483,1484])              & \textbf{1521 ([1520,1522])}     & 1530 ([1528,1531])              & \textbf{1310 ([1306,1313])}     & 1360 ([1355,1365])              \\ \hline
\end{tabular}%
}
\caption{Bootstrapped 95\% confidence intervals for the RIMSE with $A_0 = 10000, j_0 = 3, J = 7$ and $M = 1$. The number in bold indicates the best choice of $\alpha$ for each method.}
\label{ci_thresh_l10000_alpha}
\end{table}

\begin{table}[h!]
\centering
\resizebox{\textwidth}{!}{%
\begin{tabular}{l|l|l|l|l|l|l|}
\cline{2-7}
                                            & \textbf{LRT-L, $\alpha$ = 0.01} & \textbf{LRT-L, $\alpha$ = 0.05} & \textbf{LRT-I, $\alpha$ = 0.01} & \textbf{LRT-I, $\alpha$ = 0.05} & \textbf{LRT-G, $\alpha$ = 0.01} & \textbf{LRT-G, $\alpha$ = 0.05} \\ \hline
\multicolumn{1}{|l|}{\textbf{Blocks}}       & 1744 ([1742,1747])              & \textbf{1637 ([1635,1640])}     & 1553 ([1551,1555])              & \textbf{1493 ([1490,1495])}     & \textbf{1655 ([1651,1659])}     & 1770 ([1765,1776])              \\ \hline
\multicolumn{1}{|l|}{\textbf{Bumps}}        & 3500 ([3497,3503])              & \textbf{3241 ([3238,3244])}     & 3112 ([3109,3115])              & \textbf{2971 ([2968,2973])}     & \textbf{3070 ([3069,3072])}     & 3065 ([3063,3066])              \\ \hline
\multicolumn{1}{|l|}{\textbf{TriangleSine}} & \textbf{1453 ([1452,1453])}     & 1458 ([1457,1458])              & \textbf{1499 ([1498,1500])}     & 1514 ([1513,1515])              & \textbf{1269 ([1266,1272])}     & 1304 ([1299,1308])              \\ \hline
\end{tabular}%
}
\caption{Bootstrapped 95\% confidence intervals for the RIMSE with $A_0 = 10000, j_0 = 0, J = 7$ and $M = 1$. The number in bold indicates the best choice of $\alpha$ for each method.}
\label{my-label}
\end{table}

Decreasing $\alpha$ from 0.05 to 0.01, and thus making the hypothesis tests more conservative, seems only interesting for LRT-G as it decreases its RIMSE for all intensity models and both choices of $j_0$. For all other methods, choosing $\alpha = 0.01$ lead to a loss of performance on \textit{Blocks} and \textit{Bumps} whereas their RIMSE slightly decreases for \textit{TriangleSine}. Again, the effect of one parameter on the RIMSE is very specific to each intensity model.

\bibliographystyle{rss}

\bibliography{refs}

\end{document}